\documentclass{article}
\usepackage{graphicx}
\usepackage{amsmath}
\usepackage{amsfonts}
\usepackage{amssymb}

\newtheorem{conjecture}{Conjecture}

\begin{document}
\bibliographystyle{unsrt}

\title{Equilibrium properties of a  hysteresis dimer molecule from MD
simulations using two-body potentials  }
\author{Christopher G. Jesudason 
\thanks{the initial theoretical conceptualization and writing up of the program and algorithm development was done at Norwegian University of Science and Technology-NTNU, Institute of Physical Chemistry, Realfagbygget, 7491 Trondheim, Norway during a  sabbatical visit 2000-2001 financed by University of Malaya, and  hosted by Signe Kjelstrup. The method of MD used here conforms to  the periodic boundary conditions and thermostatting algorithms developed or refined by T. Ikeshoji and B. Hafskjold  which is a non-synthetic approach using very traditional integration of Newton's equations of motion for the dynamics portion. }\\
{\normalsize Chemistry Department, University of Malaya}  \\
{\normalsize  50603 Kuala Lumpur, Malaysia}}
\date{\normalsize 12 September,2005}
\maketitle
\begin{abstract}
Recent experiments indicate that  electromagnetic hysteresis behavior can be exhibited at the molecular level. Based on these indications, a simple dimeric molecule with a hysteresis-like pathway with different spatial coordinates for  formation and break-up 
  is described and a MD simulation using 2-body potentials and switches to form and break bonds is implemented to determine whether chemical reaction pathways might also exhibit analogous behavior whilst preserving conventional  thermodynamical  outcomes.  
  The results of various common thermodynamical and kinetic properties are presented, where no unusual 
  thermodynamics is observed for the  chemical reaction with  the loop pathway  which might  be interpreted as not 
  being "`time reversible invariant"' and therefore susceptible to manifesting unusual thermodynamical phenomena. This system
  may well model particles that interact via the electro-magnetic field to form  molecules, since hysteresis
  behavior is standard  and well represented in large-scale magnetic and electrical systems, especially in the 
  solid state. The potential switching  technique circumvents the problem of
computer intensive three-body calculations which makes this model
particularly suitable for numerical investigations in both equilibrium
and  nonequilibrium states where the details of a particular reaction
mechanism and its reaction coordinates   are not the focus of
attention. A new algorithm for the conservation of energy and momentum is incorporated in regions where the potentials are switched. The thermodynamical parameters determined  include the standard free energy, enthalpy and entropy, the activity coefficient ratios, the equilibrium constant and the many energy distribution functions of the molecule, all of which are compared to the Maxwell distribution function.Both the free dimer and  atom particle kinetic energy distributions  agree
fully with Maxwell-Boltzmann statistics but  the distribution for the relative kinetic energy of
 bonded atoms does not, thereby opening to question recent far-from-equilibrium theories such as MNET that make use of this presupposition in a fundamental sense. The  kinetic parameters determined include the rotational and translational diffusion
coefficients and the Arrhenius parameters. Most applications of  rotational diffusion seem to presuppose a projection of a value about the rotational axis (leading to a cosine dependence) but here it is shown that an angular dependence is also a feasible model during a first order relaxation process.  An NEMD simulation which uses a novel difference equation to test for mass conservation  is presented which shows numerically that the principle of local equilibrium (PLE) is violated and  is at best   only an approximation.    The results
 suggest  that although the reaction is microscopically loop-like, unlike all the models routinely proposed, yet the thermodynamics is entirely "`normal"'
 and yields results that do not contradict any of the known laws of thermodynamics. It is therefore postulated that  reaction dynamics involving  hysteresis mechanisms can occur in nature  and may be experimentally verifiable, although  experimental interpretations tend to construct models that avoid such mechanisms. A  revision of the concept of "`time reversibility"' to accommodate the above results is suggested. The general design of the reaction mechanism also allows for the use of conventional potentials without hysteresis and this will be the object of future investigation.  
\end{abstract}

\section{ Introduction}
\label{sec:1}
Recently, experiments have detected the presence of magnetic hysteresis behavior at the  single molecule level \cite{acs1,itrev1}; synthesis of such systems are also a hot topic of research. \cite{sm1}. Such facts  suggest that non-single-valued functions are involved in the phase trajectory of the system. A rational extension of this concept, which has profound theoretical implications is to construct a dynamical trajectory where the region of formation of the molecule does not coincide with that of its breakdown. There has been a reluctance in the past to consider such loop or hysteresis systems because of the absence of experimental evidence of hysteresis behavior at the molecular level, and because of the influence of the belief of "`time-symmetry"' invariance which discourages such a view, which lead to the construction of dynamical pathways which were both single valued  and which did not have any  loop or circular topology; a detailed mathematical examination of these common time symmetry presuppositions -so essential to physics-  has been made\cite{cgj4,cgj5} and it was shown that such views are often not warranted or incorrect. This work reports a workable model hysteresis reaction pathway  which leads to thermodynamically consistent behavior, exhibiting properties that will require new developments in reaction theory, and it also predicts the feasibility of such mechanisms in nature. It suggests a re-definition and extension of the ideas of "`time reversibility"' and "`microscopic reversibility"'  to cater for the proposed mechanism. Incidentally,  the shape of the potentials and switching mechanism used here is surprisingly similar to \emph{experimental} discussions of the charge neutralization reaction \cite{Levine8}  
\begin{equation}\label{e1_1}
 \text{K}^+ +  \text{I}^- \rightarrow \text{K} +  \text{I}  
\end{equation}
except that the discussion does not explicitly mention the crossing
over of the $\text{KI}$ and $\text{K}^+\text{I}^-$ potentials at  
short distances (high energy) due to  the 'time-reversal"' presuppositions referred too above. The existence of a cross-over would make the potential mathematically equivalent to the present treatment and  there is good reason to suppose that such processes can and should occur in electro-magnetically induced reaction pathways (such as is manifested in  charge-transfer and Harpoon mechanisms).  The  dimeric  particle  reaction  simulated may be written 
\begin{equation}\label{e18}
2\text{A}\begin{array}{*{20}c}
   {k_1 }  \\
    \rightleftarrows   \\
   {k_{ - 1} }  \\

 \end{array} \text{A}_\text{2}  
\end{equation}
where $k_1$ is the forward rate constant and $k_{-1}$ is the backward constant. The reaction simulation was  conducted at  a mean temperature which is high, $T^\ast=8.0$  well above the supercritical
regime of the $LJ$ fluid. 
There have been various attempts in modeling chemical reactions 
with different objectives in mind
\cite{Allen1,Zeiri2,Gorecki3,Stillinger4,Benjamin5,Bergsma6}. Some 
used generalized models with few details to predict  the main features
experiments might reveal \cite{Allen1} at the reaction coordinate
close to the transition state (TS), such as what might occur within a
solvent-caged reaction complex: $ \mbox{A-H} \cdots
\mbox{B}\rightleftharpoons \mbox{A}\cdots \mbox{H-B} $ . This
particular pioneering  approach  was further elaborated by Bergsma \emph{et al}
\cite{Bergsma6} in order to examine the limits of validity of TS
theory (TST) by not carrying out an ab initio study of all the possible
reactive trajectories,but by examining trajectories constrained to the
TS surface because of the limits of computing  power. An example of an
ab initio detailed chemical reaction approach with a 1000 atom system
using an assumed 3 body potential for the exchange process $ \mbox{F} +
  \mbox{F}_{2} \rightleftharpoons \mbox{F}_{2} + \mbox{F}$  is that of  
       Stillinger \emph{et al} \cite{Stillinger4} who admits that the procedure
  is 'very demanding' . At the other extreme are generalized studies
  of hypothetical schemes \cite{Gorecki3} such as the 'chemical
  reaction' $ \mbox{A} +
  \mbox{A}\rightleftharpoons \mbox{B} + \mbox{B}$ used to elucidate
  some kinetic properties. Clearly in such models, species A and B
  must represent complex systems   that can be physically
  distinguished; in chemical applications, they might represent for
  instance \emph{cis} and \emph{trans} isomers of some compound or
  they might represent mesoscopic species. Some simulations do away
  altogether with the details of molecular dynamics based on dynamical
  laws \cite{Zeiri2}, replacing them with the Ansatz  that the details
  of the interaction between individual particles are not essential in
  the study of the statistical evolution of the system. Such an
  approach would make studies attempting to correlate the details of
  the dynamics to macroscopic properties difficult or obscure, despite
  the great savings in computer time, and therefore does not suite the
  purposes at hand here. The objectives of the present study include:
\newline
  (a) designing a mechanically well defined reaction model with low
  computational demands and where the averaged motions of the dimer
  may be correlated with the macroscopic kinetic and thermodynamical
  properties and where  no anomalies must be observed in the
  macroscopic results. Such an outcome would imply  that the dynamics are
  reliable enough to be  used in other studies    
\newline
(b) introducing some degree of complexity to the dimer such as
  vibrational and rotational states for more detailed dynamical investigations
 \newline
(c) utilizing the thermodynamically consistent model (as judged by the
results of an equilibrium simulation) in nonequilibrium simulations
 \newline
Here we focus primarily on (a) above. To this end a new general algorithm (which will be discussed separately in another planned work)  was used to conserve momentum and energy.   

The following essential thermokinetic parameters will  be determined and discussed in the sections that follow: 
\begin{itemize}
\item The thermodynamic equilibrium constant through extrapolating the density to zero.
\item The activity coefficient ratio.
\item The standard Gibbs  Free energy, Enthalpy and Entropy of the reaction through extrapolation.
\item The Arrhenius activation energy and pre-exponential terms, which bears no immediate connection to the 
potential of activation in Fig.~\ref{fig:1}, and the rate constants of the forward and reverse reactions.
\item  The diverse  probability distributions for the kinetic energy about the CM (center of mass) for all the species, the internal energies of the molecule, which  \emph{do not have a Boltzmann distribution }, thereby casting into doubt many fundamental theories that assume the opposite (where is is noted that the famous Eyring Activated Complex Theory (ACT) does not consider the vibrational mode for the reaction coordinate to be active, thereby not contradicting our results)and these are compared to the Maxwell-Boltzmann distribution.
\item Self Diffusion and rotational diffusion  constants.  
\end{itemize}
  The method appears very  promising  for quantitative simulations of real systems, and  will be utilized in the years  ahead for various reaction studies.

\section{The Model}
\label{sec:2}

We  examine the  dimeric  particle  reaction given in (\ref{e18}) above 
 \[ 2\text{A}\rightleftharpoons\text{A}_{2} \]
in a range of equilibrium fluid states all well above the $LJ$ supercritical 
regime. This model resembles somewhat that of ref. \cite{Gorecki3} except that
a harmonic potential is coupled to the products to form the bond of
the dimer whenever the internuclear distance reaches  the critical
value $r_f$ between two free atoms A.
\setcounter{figure}{0}
\begin{figure}[htbp] 
\begin{center}
 \includegraphics[width=11cm]{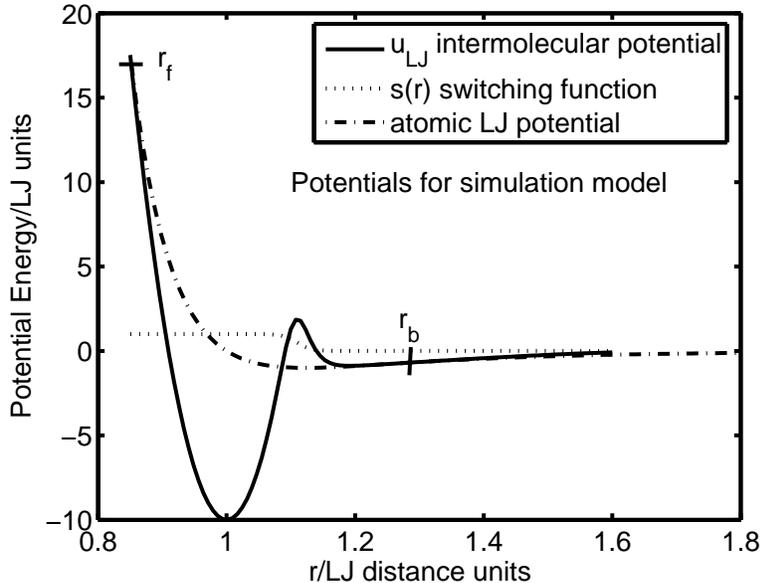}
 \caption{Potentials used for this work}\label{fig:1}
\end{center}
\end{figure}

\begin{figure}[!htbp] 

\begin{center}
 \includegraphics[width=8cm]{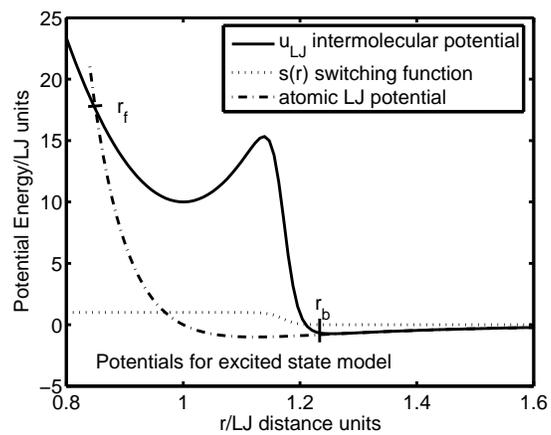}
 \caption{Potentials used for the excited molecular state }\label{fig:2}
\end{center}
\end{figure}

\begin{figure}[!htbp] 
\begin{center}
 \includegraphics[width=8cm]{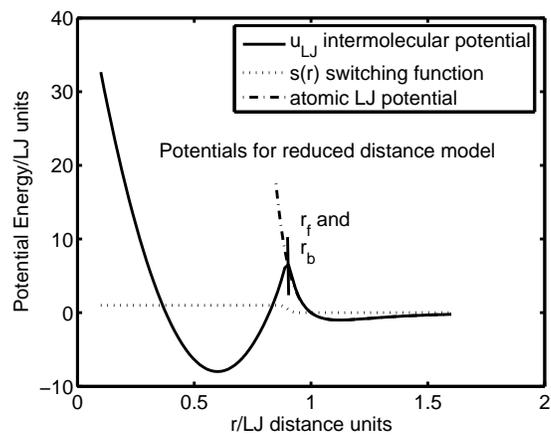}
 \caption{Potentials used for reduced distance molecular model }\label{fig:3}
\end{center}
\end{figure}

\begin{figure}[!htbp] 
\begin{center}
 \includegraphics[width=8cm]{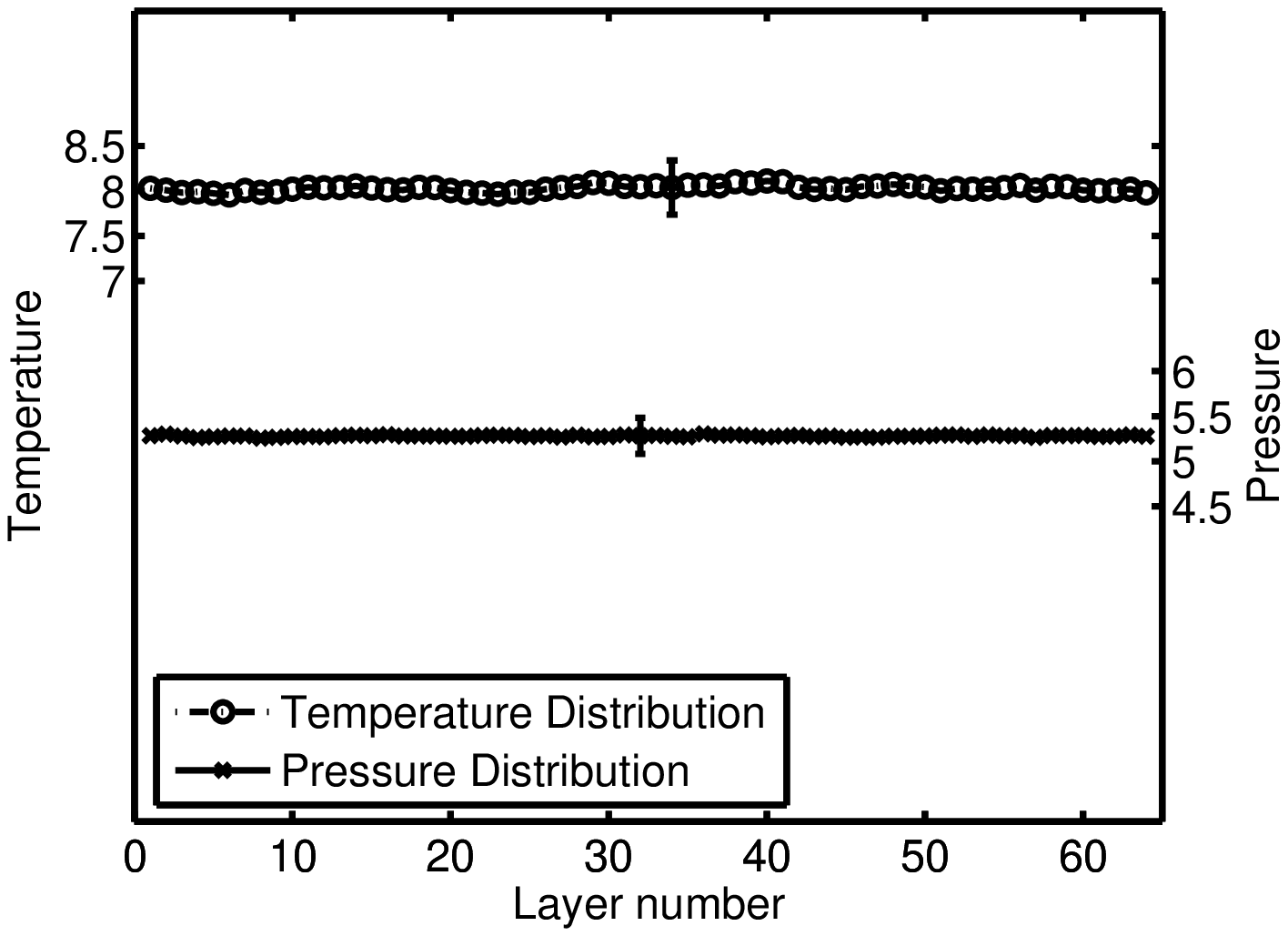}
 \caption{Pressure and temperature distribution across the MD cell }\label{fig:pt}
\end{center}
\end{figure}

In the current study, the potentials  as given  in Fig.~\ref{fig:1}
 are used, but other  configurations
are possible, as verified by direct simulation, such as the excited
 state configuration of Fig.~\ref{fig:2} and the reduced distance model 
model with the same spatial coordinates for the forward and reverse
reactions in Fig.~\ref{fig:3}. This is a typical reaction potential and it is proposed that 
a quantitative simulation of a simple dissociation reaction of a diatomic gas be attempted.
 It was found that the equilibrium exchange
rate of eqn.~\ref{e18}  was very
low at lower temperatures and changed rapidly at higher temperatures
to a saturation level   for the latter model (Fig.~\ref{fig:3}), not making it very suitable for studies where
rates of formation and breakdown of bonds must be large enough for
accurate statistics to be gained across the  MD cell over a wide range
 of density and temperature ranges; the reason for
the slow exchange is in part related to the small reaction or
collisional cross-section  of the molecule. 

The MD mechanism for bond formation and breakup is  as follows.
The free atoms A interact with
all  other particles (whether A or A$_{2}$) via
a Lennard-Jones spline potential and this type of potential has been
 described in great detail elsewhere \cite{Haf7}. An atom at a distance $r$ to another
particle possesses a mutual potential energy  $u_{LJ}$ where
\begin{align}
u_{LJ}  & =4\varepsilon\left[  \left(  \frac{\sigma}{r}\right)  ^{12}-\left(
\frac{\sigma}{r}\right)  ^{6}\right]  \text{
\ \ \ \ \ \ \ \ \ \ \ \ \ \ \ \ \ \ \ for }r\leq r_{s}\\
u_{LJ}  & =a_{ij}(r-r_{c})^{2}+b_{ij}(r-r_{c})^{3}\text{
\ \ \ \ \ \ \ \ \ \ for }r_{s}\leq r\leq r_{c}\nonumber\\
u_{LJ}  & =0\text{ \ \ \ \ \ \ \ \ \ \ \ \ \ for }r> r_{c}\nonumber
\end{align}
and 
\ where $r_{s}=(26/7)^{\frac{1}{6}}\sigma $ \cite{Haf7}. The
molecular cut-off radius $r_c$ of the spline potential is such that \ $r_{c}%
=(67/48)r_{s}$. The sum of particle diameters is $\sigma$\ and
$\varepsilon$ is the potential depth for interactions of type A-A (particle-particle)
or A-A$_{2}$  (particle-molecule) designated (1-1) or (1-2) respectively. 
 The constants $a_{ij}$\ and $b_{ij}$\ where given before
\cite{Haf7} as 
\begin{align}
a_{ij}  & =-(24192/3211)\varepsilon/r_{s}^{2}\nonumber\\
b_{ij}  & =-\left(  387072/61009\right)  \varepsilon/r_{s}^{3}%
\end{align}
The potentials for this system  is illustrated in Fig.~\ref{fig:1}. 
 Any two unbounded atoms
interact with the above $u_{LJ}$  (1-1) potential up to  distance  $r_{f}$
with energy $\varepsilon=u_{LJ}(r_{f})$ when the potential is switched
at the cross-over point  to the molecular potential given by

\begin{equation}
u(r)=u_{vib}(r)s(r)+u_{LJ}\left[  1-s(r)\right]\label{e4}
\end{equation}
for the interaction potential  between the bonded particles   constituting the
molecule where $u_{vib}(r)$ is the vibrational potential given by
eq.(\ref{eq:5}) below and the switching function $s(r)$ has the form
given by eq.(\ref{eq:6}) . LJ reduced units are used throughout this
 work unless stated otherwise by setting $\sigma \,\text{and}
 \, \varepsilon $ to unity in the above potential description.The
 relationship between normal laboratory units, that of the MD cell
 and the LJ units have been extensively tabulated and
 discussed  \cite{Haf7}and will not be repeated here . For the system simulated here with the
potentials depicted in Fig.~(\ref{fig:1}), the switching function is operative
up to  $r_b$, the distance at which the molecule ceases to exist, and where
the atoms which were part of the molecule  interact with the $(1-1)$ potential
$u_{LJ}$ like other free atoms;bonded atoms interact with other
particles , whether  bonded or free with the $u_{LJ}$ (1-2) potential. The
point $r_f$ of formation corresponds to the intersection of the
harmonic  $u_{vib}(r)$ and $u_{LJ}$ curves , and their gradients are
almost the same  at this point; by the Third dynamical law, momentum
is always conserved during the crossover despite finite changes in the
gradient, since the sudden change of the force field is between only the two particles where the Third Law applies, thereby conserving momentum. Total energy is conserved since the curves cross, and errors can
only be due to the finite time step per cycle in the Verlet leap frog
algorithm,which would cause the atoms to be defined as molecules at
distances $r<r_f$. Similarly at the point of breakup, there is a very
small ($\sim10^{-4}$ LJ units of energy) energy difference between the
LJ and molecular potentials despite using the switching function
 in the vicinity of the region to smoothen and unify the curves; the
 small energy differences at the cross-over points are less than that
 due to the normal potential cut-off at distance
 $r_c$ where the normal (unsplined) LJ potential is
 used in MD simulations. In order to overcome this problem, a new algorithm (NEWAL) was developed, the details of which will be described in another work which conserves momentum and energy at these cross-over points.  If $E_p (r)$ is the inter-particle potential (energy) and  $E_m (r)$  that for the molecule just after the crossover, the algorithm promotes the particles to a molecule and rescales the particle velocities of only the two atoms forming the bond from $\mathbf{v}_\mathbf{i}$   to $\mathbf{v'}_\mathbf{i}\;\;(i=1,2)$      where 
 $\mathbf{v'}_\mathbf{i}  = (1 + \alpha )\mathbf{v}_\mathbf{i}  + \mathbf{\beta }$  such that energy and momentum is conserved, yielding  $\mathbf{\beta } = \frac{{ - \alpha (m_1 \mathbf{v}_\mathbf{1}  + m_2 \mathbf{v}_\mathbf{2} )}}
{{(m_1  + m_2 )}}$ (for momentum conservation) and energy conservation implies that $\alpha$  is determined from the quadratic equation $\alpha ^2 qa + 2qa\alpha  - \Delta  = 0$  with $a = (\mathbf{v}_\mathbf{1}  - \mathbf{v}_\mathbf{2} )^2 $ ,$q = \frac{{m_1 m_2 }}{{2(m_1  + m_2 )}}$  and $\Delta  = (E_p  - E_m )$ . Interchanging $m$ and $p$ allows for the same equation to be used for break-up of the molecule to free particles. For the simulations,  success in real solutions for $\alpha$  for each instance of molecular formation is 99.9 \%  and 100\% for breakdown-where the $\Delta$  value in this   instance is very small ( $ \sim 1.0\times 10^{-4}$). In these simulations, we ignored the cases when there was no solution to the quadratic equation, meaning no molecules are formed at all, and the interactions are of the $(1-1)$ variety. This new algorithm coupled with shorter time step (from $0.002^\ast$ to $0.00005^\ast$ ensured excellent thermostatting, where the thermostating was carried out at the ends of the box, as is the case in most real systems. It should be noted that this much smaller time scale is not unrealistic as the temperature for this system is of the order of $20-30$ larger than the usual values chosen, and so the translational kinetic energy of the particles would scale by the same order. In this equilibrium study, the MD cell (which is a rectangular box) is divided into 128 equal orthogonal layers in the x direction,
which is of unit length in cell units. In this method of boundary conditions \cite{Haf7} , the first $64$ layers to the midpoint along the $x$ axis are a mirror reflection about the plane parallel to the other two axis passing through this $x$ axis mid-point.  The $y$ and $z$ directions
have length $1/16$ each (cell units). This shape is chosen because  non-equilibrium   simulations will 
concentrate on imposing thermal and flux gradients along the
$x-$axis,which would allow for more accurate sampling of steady state
properties about this axis \cite{cgj6}. The layers that are mirror reflections about the mid-point plane are averaged for  steady state thermodynamical properties, leading to effectively $64$ layers.   With this algorithm, with only wall thermostatting, we sample each of the layers for temperature and pressure changes, and find that the profiles are rather constant, as shown in Fig.~{\ref{fig:pt}}. The heat supply term (per unit time) are zero to within the error of fluctuation of energy. Without the algorithm, (but with the same time step increment )the center of the effective cell (layer $32$ ) would have a temperature $T^\ast$  higher than that of the thermostatted end layers by over $2$ units, and the heat supply term would be significantly negative, implying a virtual heating up of the system at the middle  due to the potential differences due to the switches at the crossover points. The pressure too would be unrealistically higher at the center of the cell, which is unphysical. The algorithm above therefore is very effective in overcoming these problems. Prior to this, each layer would be thermostatted to maintain a constant temperature and pressure profile.  
 conversion units
At regions
$r<r_{sw},s(r)\rightarrow1$ implying $u(r)\sim u_{vib}(r)$,~i.e. the
internal force field is essentially harmonic for the molecule and at
distances $r<r_{sw},u(r)\sim u_{LJ}$, so that  the particle approaches
that of the free LJ type.Concerning the mechanism for the
switching,in quantum mechanical kinetic descriptions, switch
mechanisms are frequently used for describing potential crossovers\cite{Levine8},
but from a classical viewpoint  one can suggest that the inductive LJ
forces due to the particle potential field   (with particles
having a state characterized  by  state variables $\textbf{s}_{LJ} $)
causes the internal variables at the critical distances and
energies mentioned above to switch to state $\textbf{s}_{M} $ when
another force field is activated for the atoms of the dimer
pair. State  $\textbf{s}_{M}  $ reverts
again to state $\textbf{s}_{LJ} $ at distances $r_b$. The shape of the
potentials and switching mechanism used here is surprisingly similar
to discussions of the charge neutralization reaction \cite{Levine8}  mentioned 
in (\ref{e1_1})  $ \text{K}^+ +  \text{I}^- \rightarrow \text{K} +  \text{I} $
except that the discussion does not explicitly mention the crossing
 over of the $\text{KI}$ and $\text{K}^+\text{I}^-$ potentials at  
short distances (high energy),
although there is reason to suppose that such processes may well
occur, since the KI potential curve exists at shorter distances well
before the crossover point.The following values were used  here for the
potential parameters:
\newline
(a) Current study (Fig.~\ref{fig:1})
\newline
$u_0=-10,r_0=1.0,k\sim2446$ (exact value is determined by the other
input parameters),$n=100,r_f=0.85,r_b=1.20,\mbox{and}\; r_{sw}=1.11$.
\newline
(b) Excited state model (Fig.~\ref{fig:2})
\newline
$u_0=10,r_0=1.0,k\sim2446$ (exact value is determined by the other
input parameters),$n=100,r_f=0.85,r_b=1.30,\mbox{and}\; r_{sw}=1.17$.
\newline
(c) Reduced distance model (Fig.~\ref{fig:3})
\newline
$u_0=-8,r_0=0.6,k\sim2446$ (exact value is determined by the other
input parameters),$n=100,r_f=0.90,r_b=0.90,\mbox{and}\; r_{sw}=0.90$.

The  intramolecular vibrational potential
$u_{vib}(r)$\  for a molecule is given by 
\begin{equation}
u_{vib}(r)=u_{0}+\frac{1}{2}k(r-r_{0})^{2}               \label{eq:5} %
\end{equation}
A molecule is formed when two colliding free 
particles have the potential  energy $u(r_{f})$ whenever  $r=r_{f}<r_{0},$
at the value  indicated in (a) above.  This value  can be 
defined as  the isolated 2-body activation energy of the reaction and has the value of $17.5153$ at the indicated value of $r_f$. A molecule dissociates to  
two free atoms  when the internuclear  distance exceeds  $r_{b}$ (which in this case is 
1.20).\ The  switching function $s(r)$ is defined as 
\begin{equation}
s(r)=\frac{1}{1+\left(  \frac{r}{r_{sw}}\right)  ^{n}}  \label{eq:6}
\end{equation}
where
\[\left\{\begin{array} {ll}
s(r)  & \rightarrow1\text{ \ \ \ \ \ if }r < r_{sw}\\
s(r)  & \rightarrow0\text{ \ \ \ \ \ for }r > r_{sw}
\end{array}
\right. \].
The switching function becomes effective when the distance between the
atoms  approach the value  $r_{sw}$ (see Fig.~(\ref{fig:1})). 

Some comments concerning the MD potentials are in order.
 It is generally not correct to assume that the potentials
 in Fig.~(\ref{fig:1}) represents the transition state theory (TST)
 potential surfaces; these surfaces can only be derived by computing
 the actual potential of the dimer or free atoms
 at a known internuclear distance in the presence of all the other
 species: the
 zero density limiting potentials of Fig.~(\ref{fig:1}) cannot cause
 stable molecules to exist if they were formed by  excited atoms with
 total kinetic
 in excess of the zero density activation energy since if energy is
 conserved, the 
formed molecule would (except for a finite number of kinetic  energy
values,
 depending on the model) have to dissociate again to the atomic states
 from 
which they were formed initially. There must be energy interchange at
the
 potential well of the molecular species to remove energy so as to
 prevent dissociation. This is achieved through the presence of the temperature reservoir.
 This reservoir, if it is coupled to the  system would induce 
 a system behavior whose limit at zero density would {\itshape not} be the same as
 an isolated mechanical system.  Likewise, all
 other state functions of activation (free energy, entropy, etc. ) must be
 computed as functions of all the coordinates of the particles
 involved in the interaction. The numerical magnitude
 of these functions cannot be inferred from the isolated potentials
 above. It is surmised that these are the potentials that must be used
 to determine via statistical mechanics the various system properties,
 such as the equilibrium constant and the state functions.  Here, we 
 extrapolate to zero density at fixed temperature to derive these functions, which
 cannot be inferred from mechanics only.   

\bigskip

\section{Thermodynamic results from equilibrium mixtures}
\label{sec:3}
The reacting mixture considered here were in thermodynamic equilibrium with 4096 particles. The cell  was thermostatted at the 
ends of the cell maintained at the same temperature. Typical
runs of 10 million time steps were performed  per run at each general particle density $\rho$
(where $\rho$ is determined
 as a general density irrespective of whether the particle is free or
 is part of a molecule), where the first $200,000$ steps were discarded
 so that proper equilibration could be achieved for our data
 samples. The sampling methods have been previously described
 \cite{Haf7} where sampling of all data variables were done each
 $20^{th}$ time step and where there were $100$ dump values where each dump consists typically of $5 \times 10^5$ samples which are averaged. The $100$ dump values are then averaged again to yield the standard errors of all variables.  Dynamical quantities however had to be sampled at each
 time step $\delta t^*=0.00005.$ The thermostatting method conserves momentum and registers the energy absorbed at the thermostats \cite{Tamio9}. All parameters given here are relative to LJ reduced units, sometimes  denoted by $\ast$.
\subsection{Equilibrium constants}
\label{sub:3.1}

\begin{figure}[htbp]
\begin{center}
 \includegraphics[width=10cm]{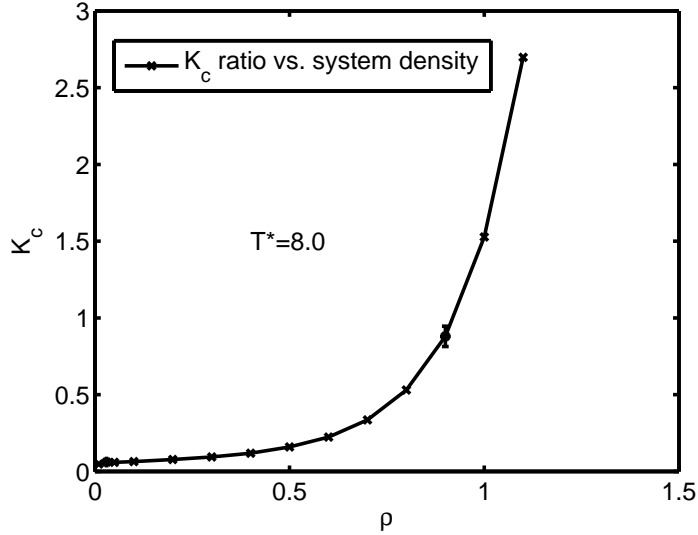}	
	\caption{Variation of  concentration ratio $K_c$ with $\rho$, the system number density at LJ temperature $T^*=8.0 $  with $sd=3$ at $\rho=.03$ and $sd=50$ at $\rho=.9$  }\label{fig:4}
	\end{center}
\end{figure}

\begin{figure} [htbp]
\begin{center}
\includegraphics[width=10cm]{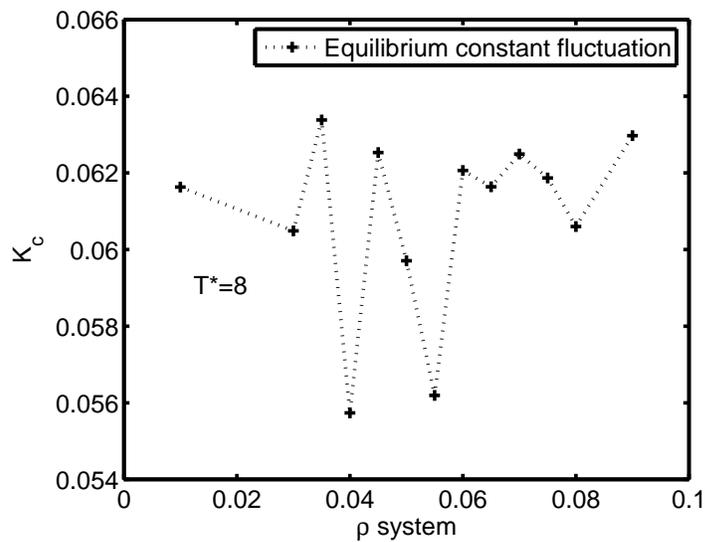}
\caption{Illustration of fluctuation of individual runs  }\label{fig:4_1}
\end{center}
\end{figure}

In order to find the thermodynamic equilibrium constant, $K_{eq}$, the
following procedure was adopted. The concentration ratio, $K_{c}$
defined as
\begin{equation} \label{eq:ec}
K_{c}=\frac{x_{\text{A}_{2}}}{x_{\text{A}}^{2}}%
\end{equation}
was determined  
as a function of average system density, $\rho$ \ where the $x$'s  
represent number density concentrations. For this and other static quantities, the temperature was set at $T^\ast =8.0$.  At very small densities, the
system becomes an 'ideal' mixture, but as mentioned previously, the
limit of the potentials cannot be the same as the isolated potentials
used in the MD calculations, since if this were the case, all the
molecules would break up, yielding a net zero value for the
equilibrium constant at the limit of zero density. As another project,
it would be of interest to determine the limit at which the
equilibrium regime breaks down in this thermostatted system,and to elucidate the theory when this occurs. There
may well be technical difficulties involved in computations of very
low density systems though .
 The plot of $K_{c}=K_{c}(\rho)$\ is shown in Fig.~(\ref{fig:4}). The accuracy of the $K_c$ values varies inversely with $\rho$, where in the captions $sd$ refers to the number of standard deviations of the standard error. At low densities, fluctuations in $K_c$ implies that any extrapolative method can be ruled out, unlike previously (when NEWAL was not devised)  when all the layers were individually thermostatted and where a  least squares fit $n$ order polynomial expansion $p(x)=\sum_{i=0}^{n}a_ix^i$ to derive the zero density limit of the concentration ratio was utilized; the value of $n$ was between $2$ to $4$.  The zero density limit $K_0$ where $K_0(T^\ast)=K_c(\rho\rightarrow 0)$is the true equilibrium constant. It is clear that in this system $K_0$ and $K_c$ in general differ significantly; it serves as a warning that in general, one cannot ignore activity coefficients in the calculation of such properties in model systems and theoretical demonstrations.  
In the present study, it was discovered that at very low densities, fluctuations are significant as shown in Fig.~(\ref{fig:4_1}) for the case of a run at $T^\ast=8.0$.
 The method used in the present case is to take the mean  value of $K_c$ for very low  $\rho$ values ranging from $0.03 \, \text{to} \, 0.09$, for about 12 values at any one temperature and to approximate this as $K_0(T^\ast)$. The fluctuations show that in this range of density, the system has "`saturated"' itself in that all the $\rho$ values yield approximately the same mean $K_c$. The results derived for $T^\ast=8.0$  are  
\begin{equation}
K_{eq}=\lim_{\rho\rightarrow0}K_{c}=0.0610\pm.002 \,\,\text{LJ units.}\label{kc}
\end{equation}
In previous studies prior to NEWAL implementation, using polynomial extrapolation, a value of $0.050 \pm .001 $ was derived. Knowing this value, we  calculate the activity coefficient ratio, $\Phi$,
for the other densities at the same temperature by  using
\begin{equation}
K_{eq}=K_{c}\frac{\gamma_{\text{A}_{2}}}{\gamma_{\text{A}}^{2}}=K_{c}\Phi
\end{equation}
The ratio of activity coefficients $\Phi$  is shown as a function of density in
\begin{figure}[htbp]
\begin{center}
 \includegraphics[width=9cm]{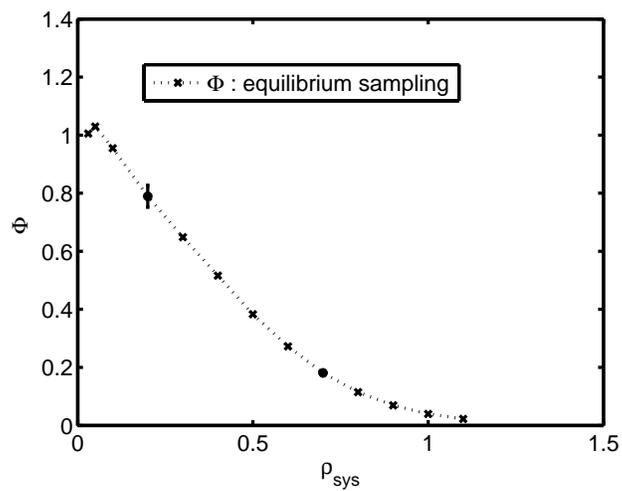}	
 \caption{Variation of $\Phi $ with $\rho$, the system number  density at LJ temperature $T^*=8.0$  }\label{fig:5_9}

	\end{center}
\end{figure}
Fig.(\ref{fig:5_9}).

It is clear from the $\Phi$ ratio that for normal densities, the equilibrium reaction mixture is highly non-ideal,which may be expected due to the large 
differences in the LJ energy well for the molecule and the atom  (see Fig.(\ref{fig:1}). It is probably a poor approximation to use ideal models for test systems in reactor design, which is often the case. Further, the above 
technique allows for the general determination of activity coefficient ratios via simulation.
 The determination of separate activity coefficients is a challenge. One real problem is the fact that molecules, in the \emph{equilibrium} state cannot exist in isolation. In mixtures, either the reaction goes to completion, or they do not react  in simple theory of mixtures. In these cases one might postulate separate ideal states for the "pure"' components, but in the present elementary case, for any one temperature, there is a finite value for $K_0$ meaning the presence of all components in a system at equilibrium. It it therefore a challenge to find a suitable model or concept to solve this problem with cycle changes. Even if a hypothetical state were defined, one must still design the route or cycle taken to the equilibrium state which consists of product and reactant species. The derivation would require  a series of very elaborate and detailed computations  and  is not attempted here since it is not immediately relevant.  The rate constant is a  defined quantity, which accords with the standard form below.  The overall
 rate of reaction $r$ may be written in terms of the
 experimentally determined forward rate ($r_1=k_1 x_A^2$) for the process 
 $2\text{A} \stackrel{k_1}{\rightarrow}\text{A}_2$  
 and backward rate ($r_{-1}=k_1 x_{A_2}$)  for the process $\text{A}_2 \stackrel{k_{-1}}{\rightarrow}2\text{A}$ as 
$r   =r_{1}-r_{-1} =k_{1}x_{\text{A}}^{2}-k_{-1}x_{\text{A}_{2}}$  %
  \. At equilibrium  $r=0,$ and so
\begin{equation}
\frac{x_{\text{A}_{2}}}{x_{\text{A}}^{2}}=\frac{k_{1}}{k_{-1}}.%
\end{equation}
The ratio of rate coefficients is the concentration ratio  $K_{c}$ where %
\begin{equation}
K_{c}=\frac{k_{1}}{k_{-1}}%
\end{equation}
To verify the above equilibrium constant independently from kinetic measurements, we can extrapolate to zero density $\rho$ the values for  $r_{1}/x_{\text{A}}^{2}=Q=k_1$ and $r_{-1}/x_{\text{A}_{2}}=R=k_{-1}$\. The rates were calculated independently from the program  by monitoring  the number of
bonds formed or broken  for each time step $\delta t^*$ and averaging this quantity over  the $10M$  time steps. Then the relevant equations are 
\begin{equation}  \label{eq:RQ}
\lim \,(\rho  \to 0)\left( {\frac{Q}
{R}} \right) = K_{eq}  = \frac{{\lim \;Q(\rho  \to 0)}}
{{\lim \;(\rho  \to 0)}} = \frac{{Q^0 }}
{{R^0 }}
\end{equation}

\begin{figure} [htbp]
\begin{center}
\includegraphics[width=10cm]{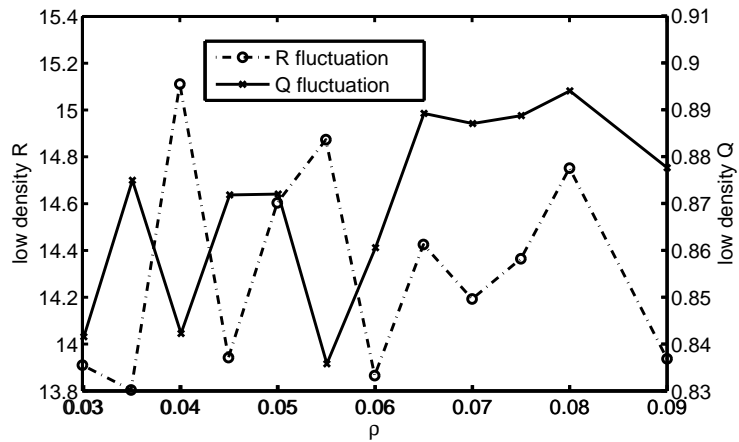}
\caption{Low density values of $Q$ and $R$ variables.  }\label{fig:7}
\end{center}
\end{figure}

The plots of $Q$ and $R$ at low densities are given in 
 Fig.(\ref{fig:7}).
 
\begin{figure} [htbp]
\begin{center}
\includegraphics[width=10cm]{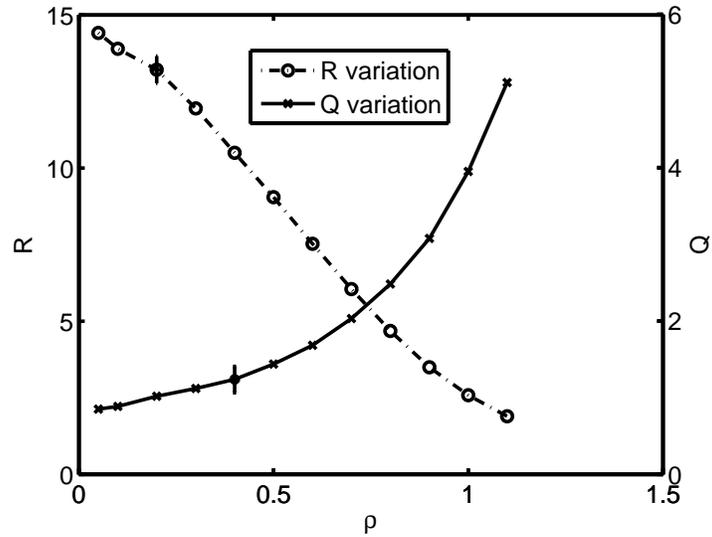}
\caption{ Variation of $Q$ and $R$ variables with density. }\label{fig:8}
\end{center}
\end{figure}

  As for the direct determination of the equilibrium constant, fluctuations imply an averaging at very low densities to derive the limits. The results with the estimated errors are 
\begin{align}
\lim_{\rho\rightarrow0}Q  & =Q^0=0.870 \pm .006 \,\,\text{L.J. units}\\
\lim_{\rho\rightarrow0}R  & =R^0=14.32 \pm .1 \,\,\text{L.J. units}.
\end{align}
It will be noticed that at very low densities, we would expect the errors due to the breakdown process to be very much higher than that due to the formation process since the number of dimers tends to a very low number.The ratio of these values gives the the true equilibrium constant directly from kinetics as  
\begin{equation}
K_{eq}(\text{kinetic})=\lim_{\rho\rightarrow0}\frac{k_{1}}{k_{-1}}=0.061 \pm.001 \,\text{L.J. units}
\end{equation}
An  excellent  agreement with the results from the previous equilibrium analysis is  found, where the method used for the determination of the equilibrium constant differs. This agreement indicates that the system is in a steady (equilibrium) state and that the simulation method is fairly coherent. The $Q$ and $R$ functions at other densities are given in Fig.(\ref{fig:8}).

\subsection{Standard states}
\label{sub:3.2}
\begin{figure}[htbp]
\begin{center}
 \includegraphics[width=10cm]{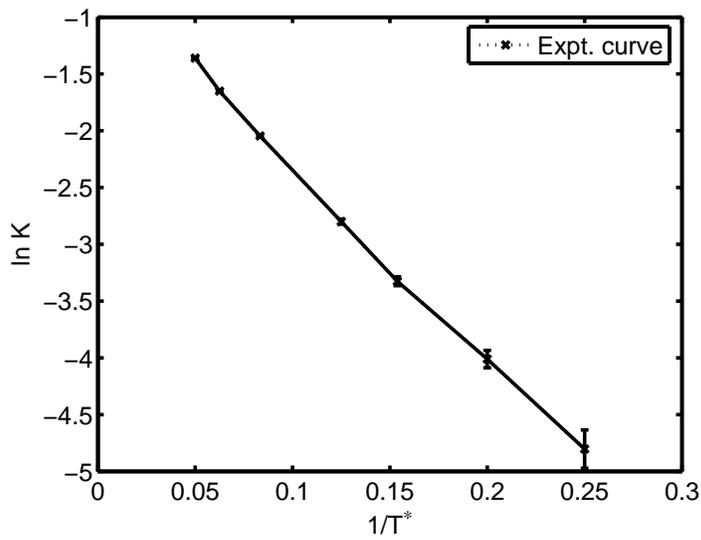}	
	\caption{Variation of equilibrium constant $K_{eq}$ with $1/T$ for fixed average system $\rho=0.70 $ with best fit line }\label{fig:9}
	\end{center}
\end{figure}

\begin{figure} [htbp]
\begin{center}
\includegraphics[width=10cm]{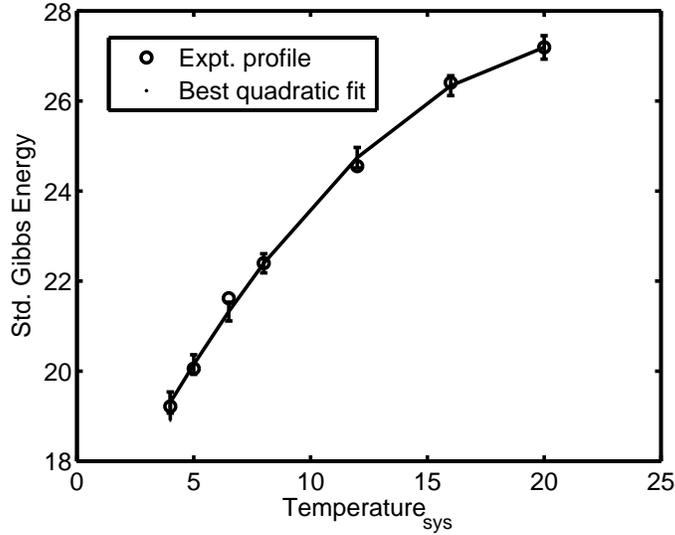}
\caption{Variation of the standard Gibbs Free Energy $\Delta G^0$ with temperature.}\label{fig:10}
\end{center}
\end{figure}

We use the form $\Delta G^0(T)=-kT\ln K_{eq}$ to determine the standard free energy  state $\Delta G^0(T)$ of the dimer  reaction.The justification is that we can choose the standard state to be  at constant pressure (of zero value)  for the standard state, so that the chemical potential standard state for each species is only a function of temperature, so that $\Delta G^0(T)$ is strictly only a function of temperature \cite[p.177-179]{ira1}. We repeat the same process as described above in  section (\ref{sub:3.1}) for $T^\ast=8$ for  different temperatures (from $T^\ast=4-20$. Each determination required at least $8$ runs at varying low densities. It was found that at  low temperatures, the fluctuations were greater, as shown in Fig~(\ref{fig:9})where the variation of $K_{eq}$ versus 1$/T$ is given.
 The curve used to determine the other standard state functions  was the Gibbs free energy curve , given in Fig.~(\ref{fig:10}). For this curve, the error bars (except for the first data set) all refer to the errors relative to the least squares fit of a quadratic curve to the simulation result. The fit is rather good.The standard entropy $\Delta S^0 (T)$is derived from the thermodynamical entity \cite[eqn. 6.34, p.182]{ira1}  
 \begin{equation}  \label{e:vh1}
 \frac{d\Delta G^0(T)}{dT}=-\Delta S^0(T)
\end{equation}
 Clearly to use (\ref{e:vh1}), we must know $\Delta G^0(T)$as a function of temperature $T$. We write therefore a simple quadratic equation with $p$ coefficients as follows
 \begin{equation}  \label{e:vh2}
 \Delta G^0(T)= p(1)T^2 + p(2)T + p(3)
\end{equation}
 
The non-linear least squares method yields  $p(1)=-0.0233441$, $p(2)=1.0531305$, $p(3)=15.46544989$ with an overall uncertainly of the free energy as approximately $\pm 0.3$. Differentiating (\ref{e:vh2}) yields the entropy 
as $\Delta S^0=-(2p(1)T +p(2))$, which is linear. The standard enthalpy $\Delta H^0$ is given at constant temperature by the entity \cite[p.183]{ira1}
\begin{equation}  \label{e:vh3}
 \Delta H^0(T)= \Delta G^0 - T\Delta S^0
\end{equation}
which therefore means that the standard enthalpy is given by $\Delta H^0 = -p(1)T^2 + p(3)$. It can be verified that this expression and that for $\Delta S^0$ recovers the quadratic (\ref{e:vh2}).

\begin{figure} [htbp]
\begin{center}
\includegraphics[width=10cm]{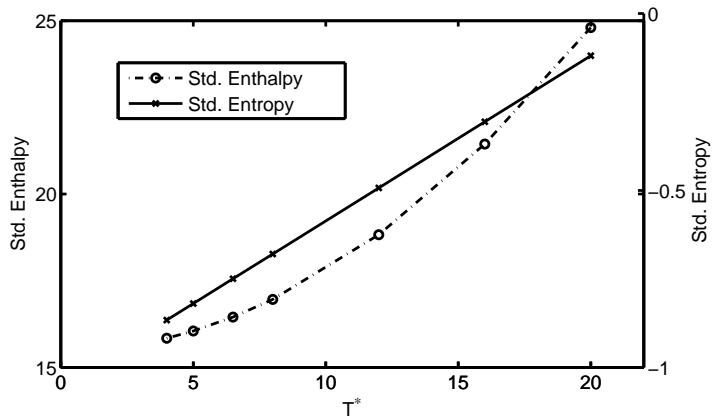}
\caption{Plot of standard enthalpy $\Delta H^0(T)$ and entropy  $\Delta S^0(T)$ from $\Delta G^0(T)$ quadratic curve fit with the temperature $T^\ast$. }\label{fig:11}
\end{center}
\end{figure}

The plots for the standard entropy and enthalpy as functions of temperature are given in Fig.~(\ref{fig:11}).

  The essential point here is that the standard entropy is negative, as it must be at moderate to low temperatures since the free particle state has a larger phase space than the corresponding dimer. It may appear counter-intuitive that the standard enthalpy is positive. It must be pointed out that at these temperatures, the particles are not trapped at the bottom of the potential well, and that the activation energy is positive, and that the internal potential energy at the point of formation of the molecule is not lost, but is converted to internal kinetic energy, leading to the break-up of the molecule. A quantitative treatment of these terms  has been attempted \cite{cg7}. It must be concluded that the simulations are able to determine the standard states without having to 
construct extremely detailed cycle diagrams; further, the simulation can also check on the correctness of the cycle diagrams used to determine standard state values.

\subsection{Activation energies}
\label{sub:3.3}
\begin{figure}[htbp]
\begin{center}
 \includegraphics[width=10cm]{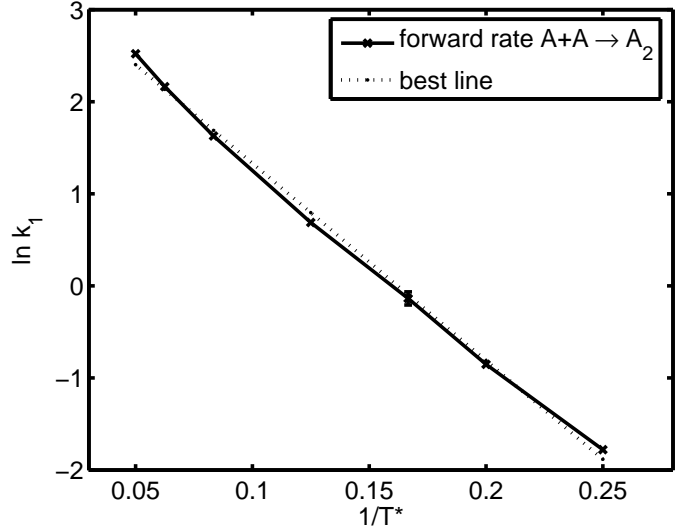}	
	\caption{Variation of natural logarithm of  forward (product forming) rate constant $k_{1}$ with reciprocal of temperature for $\rho =0.7$ } \label{fig:12}
	\end{center}
\end{figure}

\begin{figure}[htbp]
\begin{center}
 \includegraphics[width=10cm]{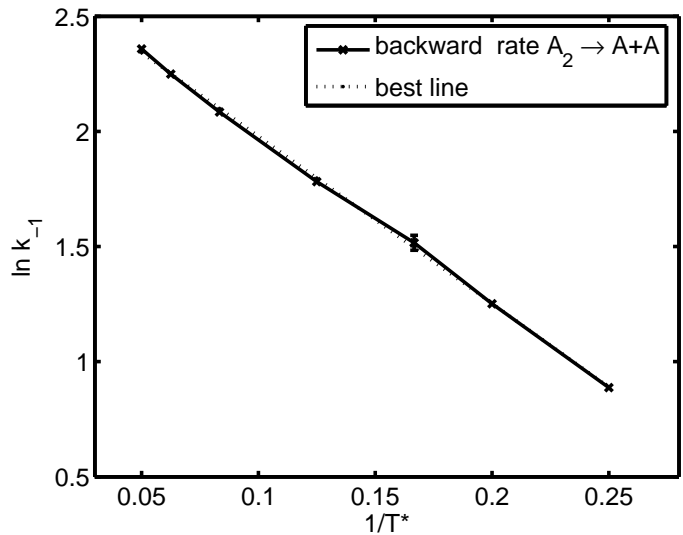}	
	\caption{Variation of natural logarithm of  backward (product disintegration) rate constant $k_{-1}$ with reciprocal of temperature for  $\rho=0.7$}\label{fig:13}
	\end{center}
\end{figure}

From the way the algorithm was constructed for molecular formation, 
the molecularity of the elementary reaction is  2  leading to a single
second-order reaction of formation, and for the dissociation of
$\mbox{A}_{2}$,a first-order reaction results since
 the molecule  can only exchange kinetic energy with all other
 particles within the system without further reactions to the
 dissociation limit. A frequently used model for the kinetic constant $k_i$  for these rates  is due to  Arrhenius,which  has the form
\begin{equation}
k_{i}=A_{i}\exp\left(  -\frac{E_{i}}{RT}\right)
\end{equation}
where the rate constant is  a function of the temperature only and where $A_i$ is ideally not temperature dependent.  
It should be noted that the Arrhenius equation is strictly valid for 2-dimensional systems where the pre-exponential factor is independent of temperature and where the exponential factor $\exp\left( -\frac{E_{i}}{RT} \right)$ represents the fraction of molecules having energy in excess of $E_{i}$ \cite{Laidler10}, where $E_i$ is usually understood to be the activation energy. The reason why this form is so durable is that the exponential term represents the fraction of excited state atoms, and this term dominates over the pre-exponential term with temperature variation, which give the impression of constant $A_i$ factor for the plots.  The rate constants for the forward $k_1$and reverse reaction $k_{-1}$  were    plotted versus $1/T$ for the  given density  of $\rho=0.7$ and was  found to be reasonably linear (Figs.~(\ref{fig:12},\ref{fig:13}), with the  
activation energies for the forward and the backward reaction rates ($E_{1}$ and $E_{-1}$ respectively) and the corresponding collision factors ($A_1$,$A_{-1}$) determined approximately as 

\begin{align*}
E_{1} & = 21.40\pm .10 \text{ LJ units}, &A_1=3.50\pm .2\text{ LJ units}\\ 
E_{-1}  & =7.26\pm .02\text{ LJ units}, &A_{-1}=2.70\pm .04\text{ LJ units} 
\end{align*}
There are two separate rate constants here, for first and second order. The second order forward rate constant    $k_{1}$ has a form given by 
\begin{equation}  \label{eq:arr}
k_1(T)=\pi b^2_{max} \left(\frac{8kT}{\pi \mu}\right)^{1/2}\exp\left(-\frac{\epsilon^\ast}{kT}\right)=A_1\exp\left(-\frac{\epsilon^\ast}{kT}\right)
\end{equation}
 Very roughly, if the mean temperature for the plot (which spans from $4$  to $20 $) is 12, then (\ref{eq:arr}) above yields for the given  value of $A_1$ $b_{max}=0.9153..$ which is reasonably  close to $0.85$, the theoretical value. However,  $\epsilon^\ast=21.40$, which is higher than $17.5153$, which is the set simulation potential value for the formation of a molecule. Since we can  expect a yet greater accuracy for the determination of $\epsilon^\ast$ as compared to $A_i$ due to the domination of the exponential terms, it may be safe to suppose that other factors contribute to the true activation energy other than what is described by simple collision theory (SCT). Future work will attempt to determine what other energy factors are implicated in $\epsilon^\ast$; currently, SCT views this energy as a pure mechanical work energy, which obtains at the molecular level. Similarly, variation of $A_i$ with various energy terms cannot be immediately ruled out. Generally, the above values do not bear a direct relationship to the isolated 2-body potentials of Fig.~(\ref{fig:1}), but nevertheless some approximate correlations are evident; $E_1$ is somewhat  close to  the isolated activation energy $17.5153$ measured from the free atomic states,and likewise  $E_{-1}$ is somewhat close to the energy difference from the bottom of the molecular potential at $-10$ to the  potential at $r_b$ which is approximately $-1$, a distance of approximately $-9$ energy units.However, for a first-order reaction, a different interpretation for energy differences obtain than from that due to SCT for instance, which is concerned with bimolecular processes; the first order interpretation is that the molecule decomposes when it overcomes an energy activation threshold, and the fraction of such molecules is reflected in the exponential term, the pre-exponential term reflecting the mechanism of  the decomposition. 

\section{Results from equilibrium  dynamical trajectory  analysis}\label{sec:4}
This section concentrates on variables which had  to be sampled at each time step of duration $\delta t=0.00005^*$ in order to compute the property of interest:the rate of reaction in the previous section above is also  based on instantaneous sampling but more  properly belongs to topics associated with equilibrium. Of importance in nonequilibrium and kinetic studies are the values of the diffusional coefficients, reaction correlation coefficients and the energy probability distributions, where if the principle of local equilibrium (PLE) obtains  imply that we may approximate  the values computed in an equilibrium simulation for those in a nonequilibrium volume element having the same state variables. Examples of these quantities (which can also gauge  the appropriateness  of the model for nonequilibrium studies) are provided. 
\subsection{Rotational diffusion constants}
\begin{figure}
\begin{center}
 \includegraphics[width=10cm]{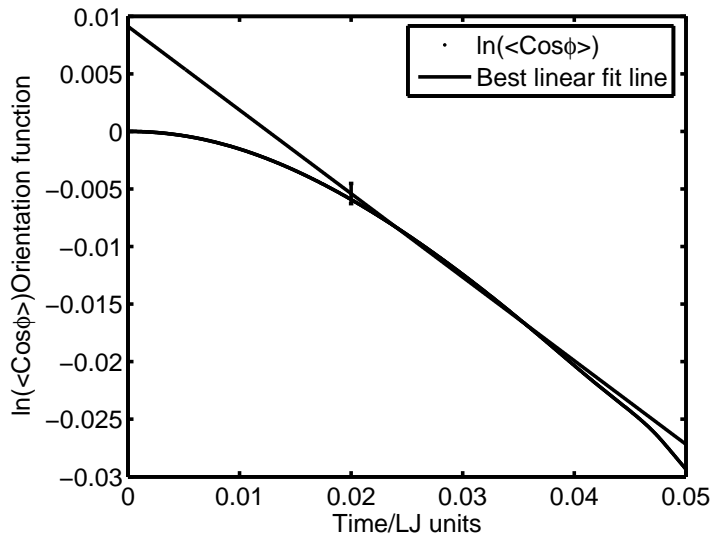}	
	\caption{Variation of natural logarithm of  $\cos \phi$ orientation function with time at $T^*=8.0 \,\text{and}\, \rho=0.7$} \label{fig:14a}
	\end{center}
\end{figure}

\begin{figure}
\begin{center}
 \includegraphics[width=10cm]{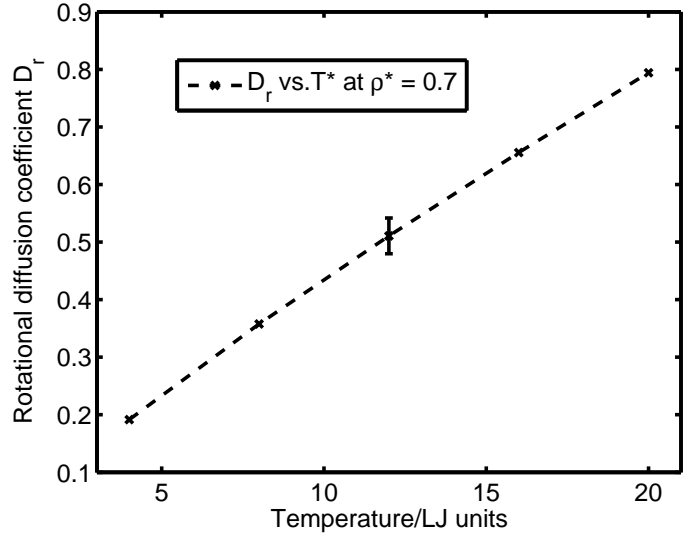}	
	\caption{Variation of $D_r$  with temperature at constant $\rho=0.7$} \label{fig:14c}
	\end{center}
\end{figure}

\begin{figure}
\begin{center}
 \includegraphics[width=10cm]{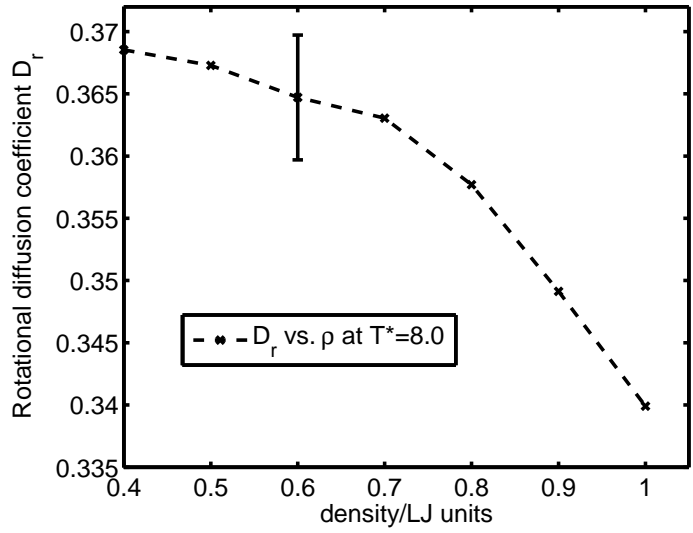}	
	\caption{Variation of $D_r$  with density at constant temperature $T^\ast=8$} \label{fig:14d}
	\end{center}
\end{figure}

Although connected in some ways to diffusion, a somewhat unconventional 'reorientation' diffusion function $\left\langle \cos \phi(t)\right\rangle $   has been defined \cite{Allen1} where $\phi(t)$ is the angle \emph{between} $\hat{\textbf{R}}(0)$, the unit internuclear distance vector  of the dimer at $t=0$, and $\hat{\textbf{R}}(t)$, the same unit vector at time $t$. Such a definition might have applications in conjunction with their being part of transform functions \cite[eqs.(17)-(20),p.211]{Allen1}, where the postulated exponential decay of this function  when acting as a kernel of the transform could  force convergence of the function being convoluted. It is found that the exponential decay assumption in $\cos(\phi(t))$is a fair but not perfect fit, perhaps  implying that  another type of theory for "`rotational diffusion"' constants may yield even  better fits with the experimental curves.We provide one such example $\left\langle  \arccos(t)\right\rangle $, an approximation to $\left\langle  \theta(t)\right\rangle$, which provides a far better fit and therefore is a candidate for  a stochastic theory of  rotational diffusion.   This then is another area for research. It must be mentioned, however, that the theory of "`rotational diffusion"' as developed by P. Debye and others \cite[p.81-84,esp eqs. 49]{debye1}  etc. makes use of "`dissipation kinetics"' where a constant torque $M$ is balanced by a inner frictional force $\zeta$ parameter, so that $M=\zeta \frac{d\theta}{dt}$ , where $\theta$ is an angular displacement. Such a theory leads to a relaxation in the distribution function $f$  by a factor $\psi(t)$ given by $\psi(t)=\exp-\frac{2kT}{\zeta}t$ so that for a particular orientation angle $\theta$, $f$ has the form $f=A\left[1+C\psi(t)\cos\theta\right]$. The mean dipole moment of the entire sample also decays with the same rate as with $\psi$. It is not immediately clear that the orientation angle must also relax according to a first order rate law. If the effect is a projection  of an orientation onto an axis,then this would correspond to the result given by Allen et al (op cit). O'Konski and Haltner \cite{haltner1} have characterized TMV (virus) by studying the birefringence relaxation rate written $\delta=\delta_o\exp(-t/\tau)$ where $\tau_o$ is the initial value of birefringence \cite[eqn 3,p.3607]{haltner1}] and the "`rotational diffusion coefficient"' $D_h$ is defined here as $D_h=1/6\tau$ with an additional  factor of $1/3$ to that of Allen. Most of these theories supposes that even at the molecular level, one can use frictional coefficients as for macroscopic systems where the retarding force is linearly proportional to some form of velocity of the system , the constant of proportionality involving the frictional coefficients \cite{broer1}. More recent studies experimental studies of rotational diffusion \cite{moog1,dorai1} assume a first order relaxation of fluorescent directed intensities of the chromophore of the molecule  with the rotational diffusion constant defined as in \cite{haltner1}.       To show that the results obtained is typical, we graph the functions as defined by Allen et. al \cite{Allen1}.  The method used here to determine $\left\langle \cos \phi(t)\right\rangle $ is to create    a table whenever a molecule is formed which maps out for each increment in the time step $i$  the value of $\cos \phi(i)$ until it disintegrates: for each $i^{th}$ time step there exists for each sampling subinterval M (M being a variable) values of $\phi(i)$ due to other molecules which have existed, and the average value for each sub-interval is computed  as
 $\left\langle \cos \phi(i)\right\rangle =\sum^M_{j=1}\cos \phi_j(i)/M$. According to Allen et al (op cit), the function decays as 
\begin{eqnarray}  \label{eq:rotdiff} \nonumber
\left\langle \cos \phi(t)\right\rangle&=&A\exp(-t/\tau_1) \,\,(A=1) \\ \nonumber
 \text{with linearized form}   & &  \\ 
    \ln\left(\left\langle \cos \phi(t)\right\rangle\right)&= -t/\tau_1 
\end{eqnarray}
 where the "`rotational diffusion"' coefficient $D_r$ is given by $D_r=\frac{1}{2\tau_1}$. 
The results of the simulation is  graphed in Figs.~(\ref{fig:14a}-\ref{fig:14d}). Fig.~(\ref{fig:14a})

  graphs the proposal found in \cite{Allen1}. It is clear that there is an initial chaotic regime, followed by a very slow  decay of approximate form $A\exp (t/\tau_r)$,($A=1$  if we measure the time from the end of the chaotic regime onwards; fitting this portion of the curve from the $400-800th$ time step to the above exponential yields   $\tau_r=1.38 \pm.02 \text{LJ units}$. A 'rotational diffusion constant' $D_r=\frac{1}{2\tau_r}$ may be defined and the  value obtained is $D_r=0.36\pm 0.01 \text{LJ units}$. The shape of the $\left\langle \cos \phi(t)\right\rangle $curve    resembles that described  in \cite{Allen1} (where the 'initial chaotic region' is mentioned)  implying a somewhat typical rotational motion, but it is clear from the figure that even in the fitting region, there is an apparent concave shape, as the tangent line makes clear. Nevertheless, for the sake of parametrization, this particular definition is used to derive the diffusion constant $D_r$ data at other regimes of varying $\rho$ (at constant temperature) in Fig.~(\ref{fig:14d}) and for  varying temperature (at constant $\rho$) as depicted in 
Fig.~(\ref{fig:14c}). In these figures, the same method of determining $D_r$ was used as for the above determination of $D_r$ at $\rho=0.7$ and $T^\ast=8$.  As with the case of rectilinear diffusion motion $D_t=BkT$, where $B$ is the density dependent mobility coefficient, which is the steady state velocity acquired per unit external force \cite[sec.14.4,eqns (2)-(11).p.464-465]{path1}, we obtain at fixed density $\rho$ a linear relationship with temperature, suggesting a similarity or isomorphous  theoretical construct in relation to rotational motion. Noting that different thermodynamical variable regimes are associated with different error margins when  determined experimentally, we also notice an approximate linear correlation with density at fixed temperature. From the rectilinear equation, this would be the case if the mobility  coefficient $B$ were inversely linearly related to the density of the medium, which is a very reasonable assumption at higher densities ($\rho^\ast=0.75-1.0$). The  figures show that the change of the diffusion constant with $\rho$  at fixed temperature is much less dramatic than with temperature at fixed $\rho$.

\begin{figure}
\begin{center}
 \includegraphics[width=10cm]{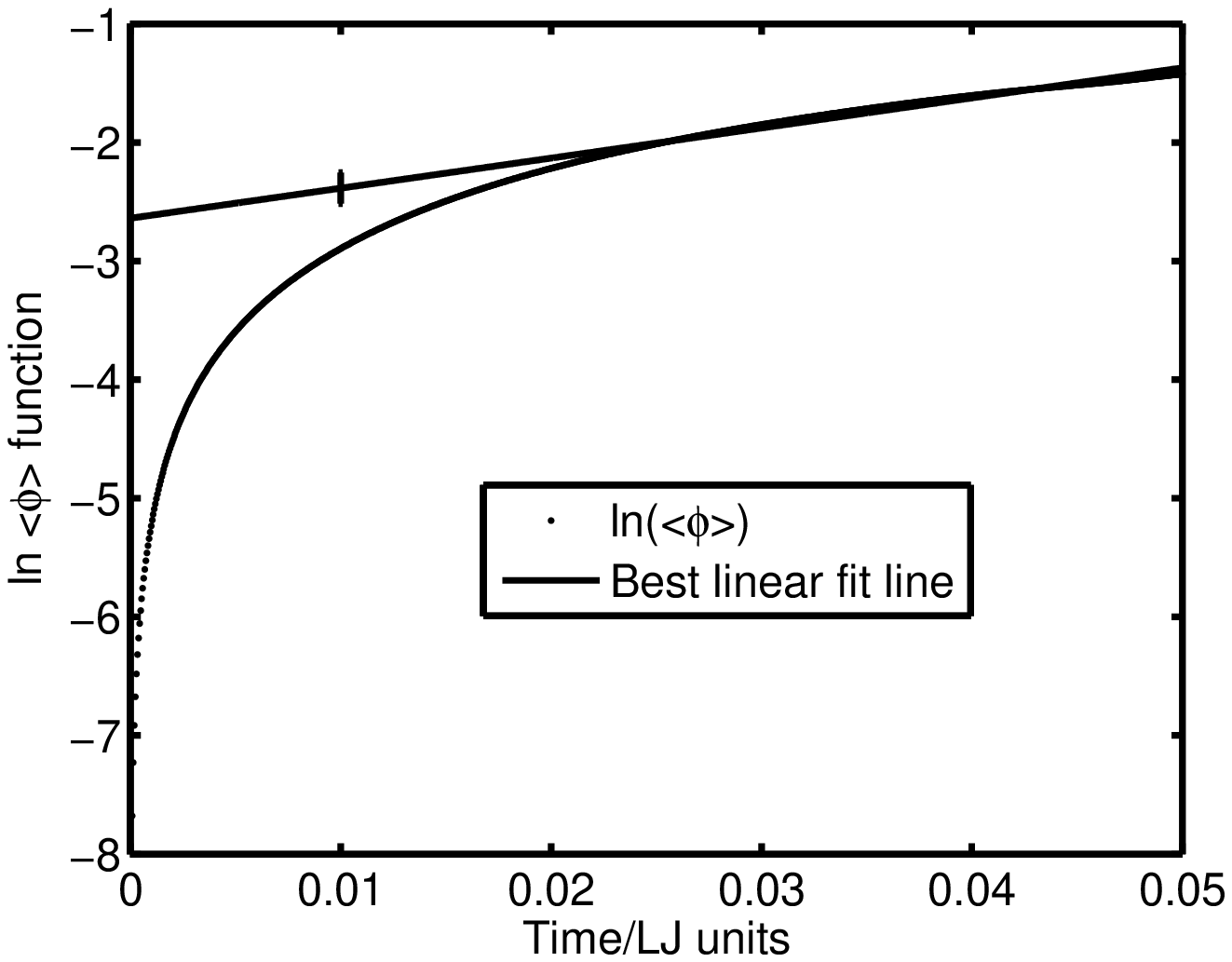}	
	\caption{Variation of natural logarithm of  $\arccos (\cos\phi)$ orientation function with time at $T^*=8.0 \,\text{and}\, \rho=0.7$} \label{fig:14b}
	\end{center}
\end{figure}

Fig.~(\ref{fig:14b}) gives a clear indication that the long-time correlation is linear concerning time and the logarithm of $\theta$, and so one can also derive a rotational diffusion theory where not a projected value, but rather the actual angular distance relaxation is a first order process. This, at any rate is what the model here depicts.
\subsection{Self-diffusion coefficients}
\begin{figure}
\begin{center}
 \includegraphics[width=10cm]{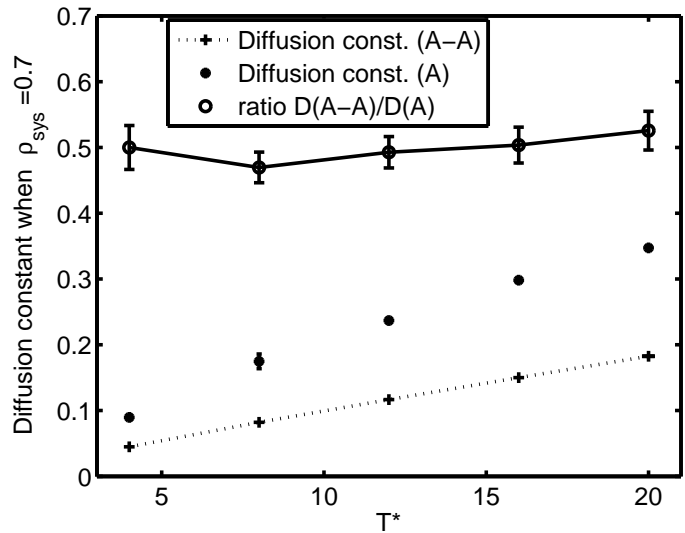}	
	\caption{Self diffusion coefficients  at varying  $T^* \,\text{and fixed} \,\rho=0.7$, where A-A refers to the dimer and A to the atom, and D denotes the self diffusion coefficient.}  \label{fig:15a}
	\end{center}
\end{figure}

\begin{figure}
\begin{center}
 \includegraphics[width=10cm]{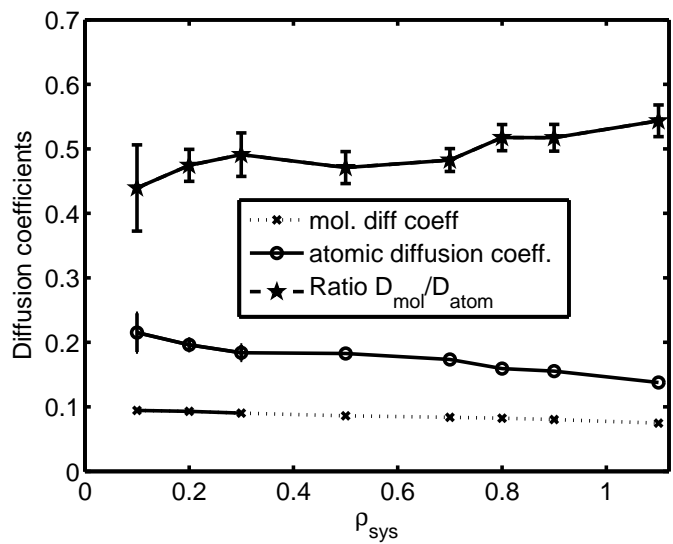}	
	\caption{Diffusion coefficients  at varying $\rho \,\text{and fixed} \,T^*=8.0 $}  \label{fig:15b}
	\end{center}
\end{figure}

In these simulations, the mean lifetime of the molecules vary in the  region of  24,000 to 2400 time steps as the corresponding temperature varies from $T=4.0$ to $T=8.0$. The  accurate determination of the three dimensional (3-D) self diffusion coefficient $D_s$ for any particle requires the determination of the  integral of the  long time limit of the velocity autocorrelation function, or the equivalent Einstein expression of the mean square displacement at infinite time with respective forms 
\begin{eqnarray} 
	D_s&=&\frac{1}{3}\int^{\infty}_{0}dt \left\langle \label{eq:ds1} \left.\mathbf{v}_i(t)\cdot\mathbf{v}_i(0)\right.\right\rangle\\\nonumber
	&\text{and}& \\ 
	2tD_s&=&\frac{1}{3}\left\langle \left|\mathbf{r}_i(t) - \mathbf{r}_i(0)\right|^2\right\rangle \,(t\rightarrow\infty)\label{eq:ds2}
\end{eqnarray}
respectively. We overcome the  infinite time problem  here by  determining the diffusion coefficient according to (\ref{eq:ds2}) at the time of breakup $t_{br,i}$ of  molecule $i$ (where the time is $0$ when the molecule is formed), thus allowing for the maximum time possible before $D_{s,m,i}$ is computed (where $m$ refers to the dimer).Likewise, we can monitor the time spent as a free particle of any labeled atomic species (j),  
and determine the self diffusion coefficient $D_{s,a,j}$ (where $a$ refers to the atomic state). The molecular self diffusion coefficient is the average of all molecules  determined during the dump interval, and lastly the 100 dump values for the entire run is averaged to provide an estimate of uncertainty. Similarly, a labeled particle is used to determine the atomic diffusion coefficient based on the time spent as a free, non-bonded particle.   The results for this supercritical fluid are given in Figs.~(\ref{fig:15a}-\ref{fig:15b}). 
The curves in Fig.~(\ref{fig:15a})appear very linear, verifying the formula $D_s=BkT$, according to previously developed theories (\cite{Levesq11}, eq.(49) ) especially at lower temperatures. The ratio of molecular to atomic diffusion constant   is relatively close to  $0.50$ everywhere. The mass of the molecule is twice that of the atom and approximately twice the diameter, leading to this approximate  ratio.The actual theoretical prediction due to size, energy interaction and mass effects is not well developed, and no extensive data are available for even non-reacting systems. The reactive system here depicts values of the diffusion coefficient which is does not differ significantly for systems which do not react. In one study \cite[p.2044 Table V]{naka1}of solute diffusion in a solvent, where interactions are solvent-solvent (1-1) and solvent-solute (1-2) only, (i.e. no (2-2) interactions) the $L_2$ system  has the following Lennard-Jones parameters $\frac{m_2}{m_1}=2;\frac{\epsilon_{22}}{\epsilon_{11}}=4;\frac{\sigma_{22}}{\sigma_{11}}=2$ leading to the diffusion  coefficients $D_1=0.063$ and $D_2=0.017$ (accuracy not specified) and for the $S_2$ system, the Lennard-Jones parameters $\frac{m_2}{m_1}=\frac{1}{2};\frac{\epsilon_{22}}{\epsilon_{11}}=\frac{1}{4};\frac{\sigma_{22}}{\sigma_{11}}=\frac{1}{2}$ lead to the diffusion coefficients  $D_1=0.082$ and $D_2=0.190$. for the same mass ratio, the diffusion constant ratios vary from $0.27$ to $0.43$ for very different and extreme $\epsilon - \sigma$ combinations where the variation with temperature is not significant for these ratios based on the scanty  information of the graphs drawn; here $\epsilon=1,\,\text{and} \sigma=1$ throughout. These ratios are not too different from the ones reported here.   The variation of the diffusion constant with density is much less dramatic than for the temperature according to Fig.~(\ref{fig:15b}) with a slight decline in diffusion constants with increasing density, as is to be expected as the mobility would decrease.The errors appear large because the variation of the coefficients with varying density is relatively slight for fixed temperature.
\subsection{Energy distribution histograms and non equilibrium results}
It is of interest to compare the theoretical Maxwell distribution of the species to the
 distributions  derived from simulation since fundamental deductions can be made. We also produce more results for a non-equilibrium simulation with a novel difference equation which can be used to check for conservation of matter to determine whether the principle of local equilibrium is indeed a principle or merely a very good approximation for describing general  thermodynamical systems (whether reversible or not).
 \subsubsection{Probability histograms}
\begin{figure}
\begin{center}
 \includegraphics[width=10cm]{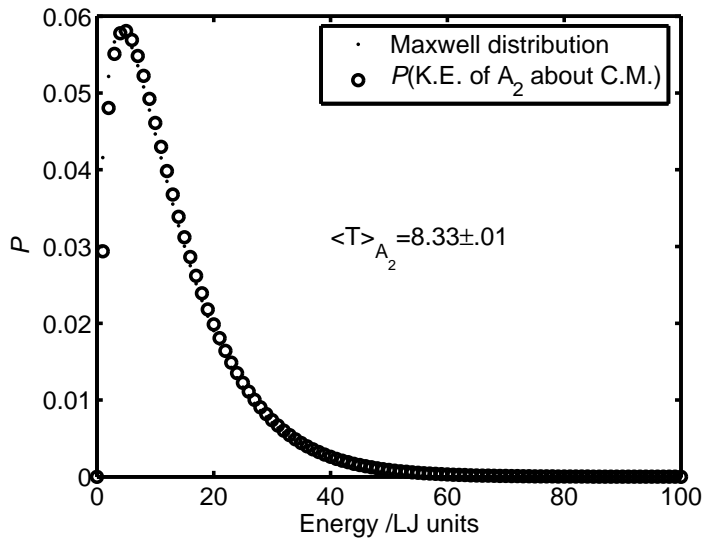}
	\caption{$P$ functions for translational kinetic  energy of $\text{A}_2$ about center of mass at system  temperature set at $T^*=8.0 \, \text{and} \, \rho=0.7$, with apparent temperature of molecule indicated. }\label{fig:16}
	\end{center}
\end{figure}

\begin{figure}
\begin{center}
 \includegraphics[width=10cm]{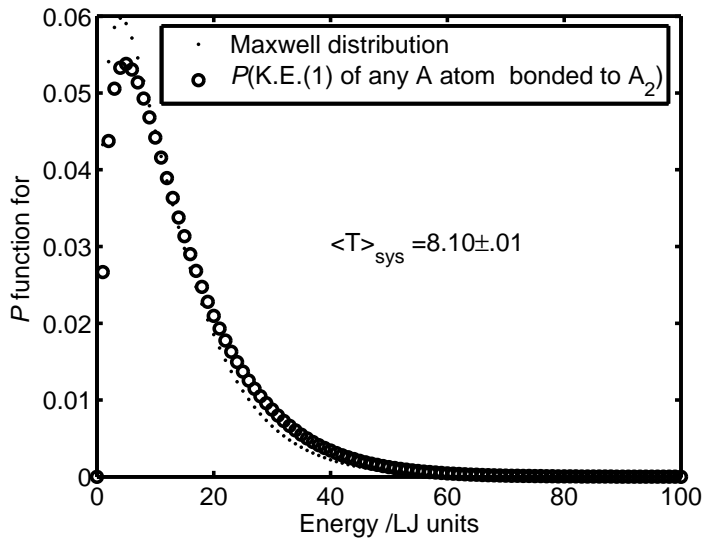}
 	\caption{$P$ functions for kinetic energy of any atom A bonded to  $\text{A}_2$ at  system temperature set at $T^*=8.0 \,\text{and}\, \rho=0.7$ with system temperature indicated.}\label{fig:17}
	\end{center}
\end{figure}

\begin{figure}
\begin{center}
 \includegraphics[width=10cm]{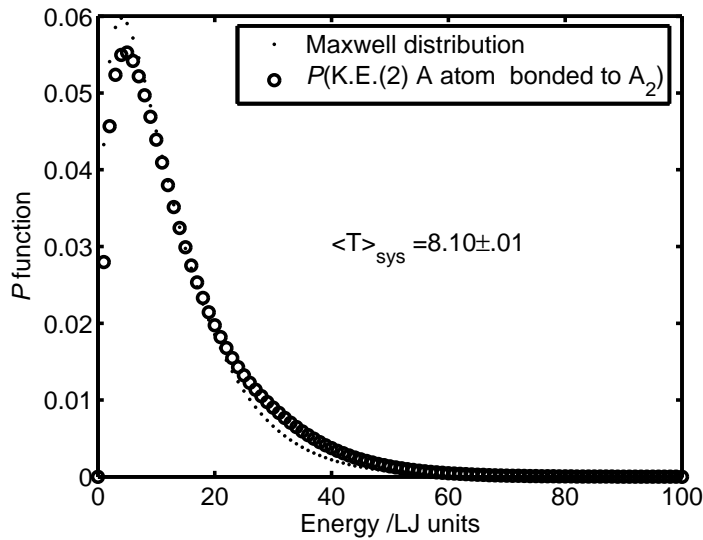}	
	\caption{$P$ functions for average kinetic  energy of atom A by K.E.(2) method  at system  temperature set at $T^*=8.0 \,\text{and} \,\rho=0.7$ with total system temperature indicated.}\label{fig:18}
	\end{center}
\end{figure}

\begin{figure}
\begin{center}
 \includegraphics[width=10cm]{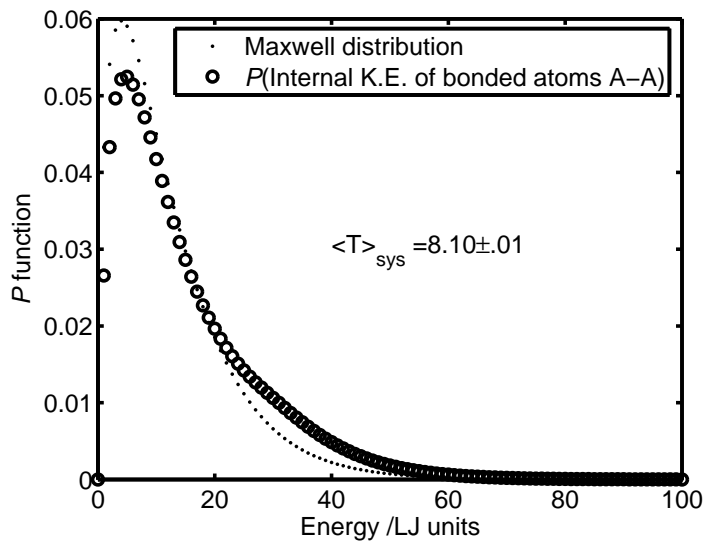}	
 	\caption{$P$ functions for total internal  kinetic energy (IKE) of the two bonded A atoms about the internuclear axis at system temperature set at $T^*=8.0 \,\text{and} \,\rho=0.7$ with system temperature as indicated.}\label{fig:19}
	\end{center}
\end{figure}

\begin{figure}
\begin{center}
 \includegraphics[width=10cm]{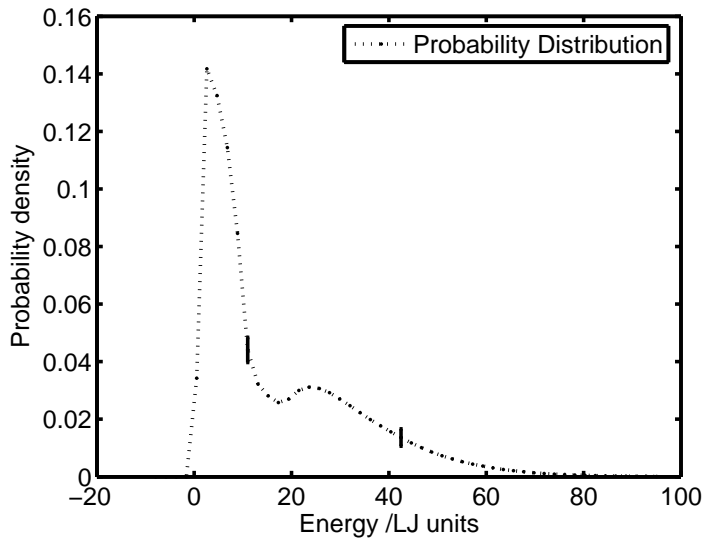}	
 \caption{The total internal energy coordinate (TIEC) distribution as given in the text. The error bars are for 100 standard error units.}\label{fig:22}
	\end{center}
\end{figure}

These are provided in Figs.~(\ref{fig:16}-\ref{fig:22}) for the translational kinetic energies of the different species probed as well as the total internal energy  of the dimer. These distributions are plotted together with the Maxwell distribution relative to the apparent temperature determined from (\ref{e:25}).  The  comparisons  provide clues to the following:
\begin{itemize}
	\item Shape   of the probability function $P$ could perhaps  be used to determine whether the  assumptions used in theories is reasonable or not.The shape  even  for this  equilibrium system is not always Gaussian, and so there is no reason to assume \emph{a priori} that non equilibrium systems must conform to a Gaussian distribution where certain internal variable are concerned. 
	\item Providing a rationale for extending the theory of equipartition in an equilibrium system where the 
temperature relative to a particular kinetic energy coordinate is not the same as for the total system temperature determined from standard equipartition. Such a possibility  seems to be  supported by the evidence below.
\end{itemize}

\begin{figure}
\begin{center}
 \includegraphics[width=10cm]{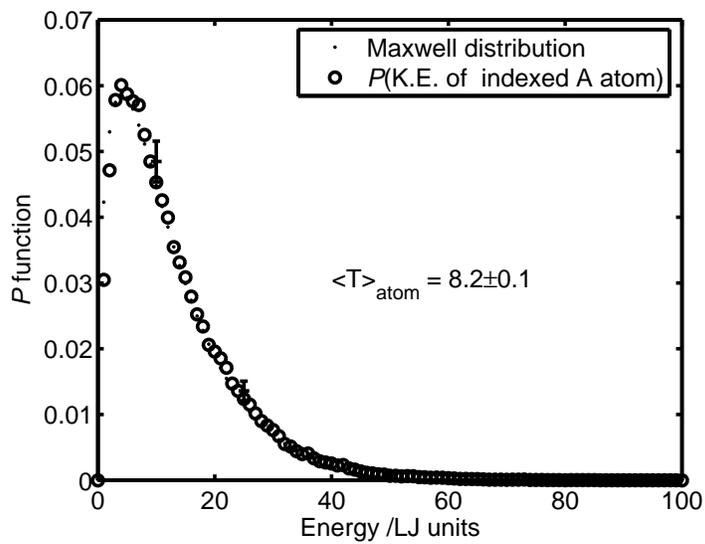}	
 \caption{$P$ functions for kinetic energy of fixed indexed atom A which either is  bonded to some  $\text{A}_2$ dimer or not at  system temperature set at $T^*=8.0 \,\text{and}\, \rho=0.7$ with apparent temperature of atom indicated.The uncertainty here is  3 standard error units. }\label{fig:20}
	\end{center}
\end{figure}

\begin{figure}
\begin{center}
 \includegraphics[width=10cm]{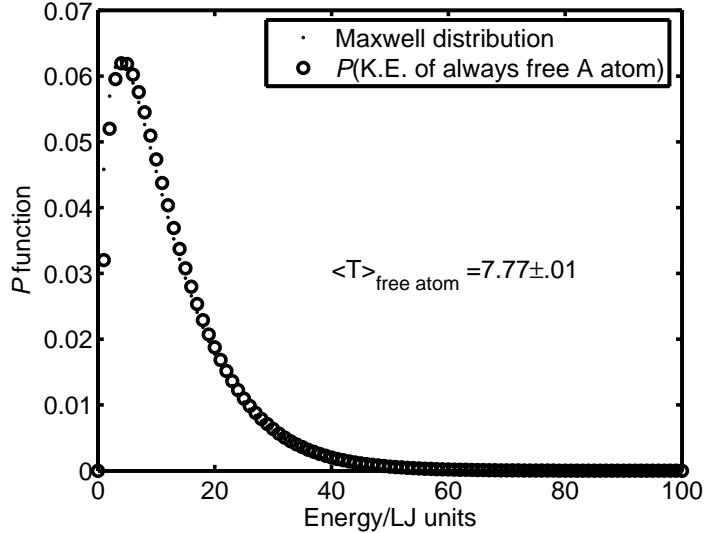}	
 \caption{$P$ functions for kinetic energy of free (unbonded) random atom  at system temperature set at $T^*=8.0\, \text{and} \,\rho=0.7$ with  apparent temperature of random atom indicated }\label{fig:21}
	\end{center}
\end{figure}

 For a given Hamiltonian $\mathcal{H}$ weakly coupled to a heat bath where 
 
\begin{equation} \label{e:25a}
	\mathcal{H} =\sum_{i=1}^{N}  \frac{p_{i}^2}{2m_i}+ \mathcal{V}(r_1,r_2,\ldots r_n)
\end{equation}

where $\mathcal{V}$ is the position variable $\mathbf{r}$ dependent potential, the probability density function per unit area of phase space $(p,q)$ is
\begin{equation} \label{e:25b}
\mathcal{P}(\mathbf{p,q})=\frac{\exp-\beta\mathcal{H}}{\mathcal{Z} }
\end{equation}
where the partition function $\mathcal{Z}$ has the form 
	\[\mathcal{Z}=\frac{\int e^{-\beta\mathcal{H}}\mathbf{}\mathbf{dpdr}}{N!}.
\]
The separability of the Hamiltonian above for  the momentum $\mathbf{p}$ and position variables $\mathbf{r}$ which is of the same form as our chemical system Hamiltonian leads for large $N$ to the exact result (in 3 dimensional systems)   (usual laboratory units) 
\begin{equation} \label{e:25}
	N\left(\frac{3kT}{2}\right)=\overline{\sum_ip_i^2/(2m_i)}
\end{equation}
which is the method used to determine the system temperature here. The momentum coordinates $p_i$ refer to all atomic species, whether bonded or not. The Gibbs postulate can be directly tested for the chemical reaction system. If this postulate is valid for loop-like hysteresis systems, then  the time trajectory of any indexed particle $I$ must also yield, when averaged over a very long time the result (in 3-D)  $3kT/2=\overline{p^2_I/(2m_I)}$ whether the particle is bonded or not over the trajectory equally weighted for all the states that it traverses. 
Integrating the $\mathcal{P}$ function in (\ref{e:25b})above over all equal energy values,  the Maxwellian probability density function results, and is given  per unit energy increment by
\begin{equation}\label{e:31}
	P=2\pi\left(\frac{1}{\pi}kT \right)^{3/2}\epsilon^{1/2}\exp-(\frac {\epsilon}{kT})
 \end{equation}
 Eq.(\ref{e:31}) is the standard form used for the absolute velocity distribution function since the energy $\epsilon\propto v^2$ for velocity $v$. An apparent temperature parameter $<T>_X$ is computed here for some species $X$ and is defined  such that $\frac{3<T>_X}{2}=\left\langle \frac{p_X^2}{2m_X}\right\rangle $ where $m_X$ is the mass of species $X$ and $p_X$ is its momentum variable. This parameter is clearly not well defined as a temperature if it does not obey the equipartition result above for the obvious reasons connected to conjugate transforms.In statistical thermodynamics, the total system Hamiltonian ${\cal H} = \sum\nolimits_{i = 1}^m {p_i^2 } /2m + \sum\nolimits_{i < j} {V(r_i  - r_j )} $ leads to the  density-in-phase  having form $
\rho ({\bf p},{\bf q}) \propto \exp [ - {\cal H}({\bf p},{\bf q})/kT
$ and so for systems with separable coordinates, each kinetic energy coordinate $
E_{k,i}  = p_i^2 /2m$  and potential form $
V(|r_i  - r_j |)
$ will have the above Boltzmann distribution. However, the "internal coordinates" during a chemical reaction or other process refer for example to an artificial aggregation such as the center of mass (C.M.)  velocity and position  for particles $k,l$ 
forming a molecule which is not permanent e.g. $
{\bf P}_j  = {\bf p}_k  + {\bf p}_l \;(k \ne l)
$ \,\(
{\bf R}_j  = \frac{1}{{m_k  + m_l }}\left( {{\bf r}_k  + {\bf r}_l } \right)
\) need not have Boltzmannized distributions.  Permanent aggregated states can be expressed in terms of canonical transformations \({\bf Q = Q(p,q),}\;{\bf P = (p,q)}\)\cite[Chap. VII]{cal1}and the new Hamiltonian that results must by ensemble theory be subjected to the density distribution described above. But for systems which are described by "`internal"' coordinates of a non-permanent nature ( in the sense that the forces between the particles cease when the molecule decomposes) and  which does not refer to the system Hamiltonian, no general theory exists, and no presuppositions can be made to  regarding its density distribution. Nevertheless, theories purporting to be  fundamental  have been created that {\itshape assumes} the Gaussian density for internal variables to be true \cite{rubi2,rubi3} without clear qualification concerning the situation when this condition obtains. A clear-cut counterexample will be provided which therefore opens to question the aforementioned theory. Furthermore, the principle of local equilibrium has been proposed as essential \cite{rubi2} for these new theories, and another counter-example to this is also provided, this time from a non-equilibrium simulation. In other words, basic simulation is able to determine the veracity of theories, and in particular, the hysteresis system described here does not support the novel theoretical developments in "`mesoscopic"' level thermodynamics. The total internal energy coordinate (TIEC) and the internal kinetic energy coordinate (IKE) are not Gaussian distributions  for equilibrium systems according to the simulation result discussed below. Of great theoretical interest is that for cases of non-permanent coordinates, some types of distributions are essentially Bolztmannized, others are not. It would be of great significance and interest to provide criteria which can predict when a Boltzmann distribution can be expected. The  apparent temperature parameter $<T>_X$  may well qualify  as a temperature in an extended equipartition  scheme if there is  agreement with the Maxwellian distribution  \emph{even if this temperature does not correspond to the unique system temperature} $<T>_{sys}$. Here the degree of agreement with the Maxwell distribution is either very good (in some cases), or rather bad. It would be of great theoretical interest if some form of relationship between the apparent  temperatures could be made on the basis of internal energetics.  The uncertainly (unless stated otherwise) is of the order as given in  the error bars of  Fig.(\ref{fig:22})which is at 100 standard error units and which would not feature in any figure where errors are typically quoted at 3 standard error units. This figure corresponds to the TIEC distribution.  The errors in the temperature are are given  in Figs.(\ref{fig:16}-\ref{fig:21}). Fig.(\ref{fig:16}) shows that the center of mass (C.M.) kinetic energy follows quite accurately a Maxwellian $P$ function with a temperature parameter  \emph{higher} ($T^*=8.33$ rather than $T^*8.0$ )  than the system  temperature. The fact that  the shape is Maxwellian at the indicated  temperature parameter does seem to imply that  theories may be be developed \emph{within}  an equilibrium system with  different coexisting temperatures provided that these parameters require that  a Maxwellian form regarding shape prevails, and after that stage one perhaps  might also be able to propose generalizations to temperature not requiring a Maxwellian distribution; but a proper theory would have to begin from first principles which can subsume without contradiction the previous axiomatics, including the Zeroth Law. Another inference is that the temperatures have definite values (or limits), since the degree of scattering is relatively low;hence one might expect some type of stochastic averaging which yields exact values (limits).  The other important scientific question is the explanation of the shift of "`temperature"' $<T>_X$  for such Boltzmann distributions for non-permanent aggregates.
An atom bonded to a molecule does not have a clear Maxwellian shape, as is evident from Figs.(\ref{fig:17}-\ref{fig:18}) since there is interference from the internuclear potentials. The graph in Fig.(\ref{fig:17}) computes the absolute kinetic energy (K.E.(1))of the particle with respect to the MD cell or AKE , whereas  Fig.(\ref{fig:18}) refers to half the relative kinetic energy and half the translational kinetic energy about the C.M.  of  the  bonded pair,where the relative kinetic energy  $\epsilon_{k.e. rel.}$, is written as $\epsilon_{k.e. rel.}= \frac{1}{2}\mu(\mathbf{\dot{r}}_1-\mathbf{\dot{r}}_2)^2= \frac{1}{2}\mu {\bf \dot r}^2 $ for any two bonded atoms 1 and 2, where the reduced mass $\mu$ is given as $\frac{1}{\mu}=\frac{1}{m_1}+ \frac{1}{m_2}$ and where the intermolecular axis vector is ${\bf r=r_1-r_2}$. The total internal kinetic energy IKE is also defined as the relative kinetic energy of a bonded pair, given as $\epsilon_{k.e. rel.}$ as above. The AKE averages $\frac{1}{2}\frac{1}{2}(\mathbf{v_1}-\mathbf{v_2})^2=\frac{1}{4}\left\{\mathbf{v_1}^2+\mathbf{v_2}^2-\mathbf{v_1}.\mathbf{v_2} \right\}$ whereas the kinetic energy about the C.M. ((KCM) averages the expression $\frac{1}{2}.2\frac{(\mathbf{v_1}+\mathbf{v_2})^2}{2^2}=\frac{1}{4}\left\{\mathbf{v_1}^2+\mathbf{v_2}^2+\mathbf{v_1}.\mathbf{v_2} \right\}$. Adding these expressions and then dividing by $2$ would lead to convergence of the result to that for AKE, which is what is presented in Fig.~(\ref{fig:18}) as K.E.(2), which is almost the same graph as for Fig.~(\ref{fig:17}).The reason for this computation was to check for consistency  of result for the two different sampling techniques. 

The IKE distribution, that of an internal coordinate, is clearly non-Gaussian, as depicted in Fig.~(\ref{fig:19}). This result is not consistent with the assumptions of mesoscopic non equilibrium thermodynamics.  \cite{rubi2,rubi3}.

TIEC defined above  refers essentially to the vibrational and rotational  kinetic energy of the molecule $E_{tiec}$ since the translational
kinetic energy about the C.M. has been factored away where 
\begin{equation}\label{e:29}
E_{tiec}  = V(|{\bf r}_{\bf i}  - {\bf r}_{\bf j} |) + \frac{{\mu {\bf \dot r}^2 }}{2}
\end{equation}
where $V(|{\bf r}_{\bf i}  - {\bf r}_{\bf j} |)=u_{0}+\frac{1}{2}k(r-r_{0})^{2}$   . Hence the intermolecular potential would play an important part in determining the motion along the internuclear axis, with the environmental potential due to other particles playing a moderating  role by introducing stochasticity to an otherwise plainly mechanical system.  The probability of occurrence of a state is proportional to the time spent at any configuration, and with a harmonic potential,  most of the time spent will be at the turning points in simple harmonic motion:in the molecular potential used there is a 'dissociation hump' just prior to the dissociation limit, leading to a departure from the Maxwell distribution;other reasons for departure form the distribution include the dissociation itself, precluding higher energy states from being accessed. It is clear that the distribution in Fig.~(\ref{fig:22})  is non-Maxwellian and accords well  with the shape of  molecular potential energy function , with its humped potential near the distance of dissociation. This model has been used as a classic description of equipartition.  If the particles were bonded permanently, this quantity would have a canonical distribution, which it  clearly does not because bonds are formed and broken at a rate that precludes adjustment to a Gaussian probability factor. This distribution , which also refers to an internal coordinate for total internal molecular energy, is not consistent with some recent non-equilibrium theories \cite{rubi2,rubi3}.

 Noting that the accuracy of the single particle is reduced by a factor of $\approx 4000$ (the number of particles in this simulation), we find that the Gibbs postulate seems to be verified in terms of the shape of the $P$ function (which appears Maxwellian) as well as the computed value of the temperature with the error estimated as $\pm0.1$ by studying an atom of fixed label (no. 29) as it forms and breaks bonds with neighboring molecules, as shown in Fig.~(\ref{fig:20}) Clearly the time average of dynamical properties for this particle would equal the ensemble average. We notice that the reduced accuracy of the sampling is reflected in  the greater scatter of the $P$ function  points.   

Finally,since the molecular $P$ function has been mentioned,it would be interesting to compare it to the case of a random, but always free A particle which is given in Fig.(\ref{fig:21}), where  the determined  temperature is slightly \textit{lower}, (to within the error limits) than the system temperature, and where the shape of the $P$ curve is Maxwellian. This particular species type cannot fulfill the Gibbs postulate because its trajectory is confined to those areas where there is no molecular formation, and so its time averaged properties like the temperature need not necessarily equal that for the system as a whole as determined from the equipartition principle.  We can conclude that the  energy subsystems that can be chosen for devising a theory of unequal temperature distributions in an equilibrium system which all have a Maxwellian probability profile include at least the following candidates:

\begin{itemize}
	\item Translational k.e. about C.M. for $\text{A}_2$
	\item Fixed indexed k.e. of particle A (in both free and bonded state)
	\nopagebreak
	\item Random, always  unbonded k.e. of particle A  
\end{itemize}
The following is suggested as a result the above observations.
\begin{conjecture}
If the random forces are external to the system, and they all have the same force law when acting on the particles of the system
which may be different from the force law for  internal forces acting on the particles of the same  system, then the kinetic energy of the C.M. would have a   probability distribution that is Maxwellian. 
\end{conjecture}
The above conjecture is weak and must be strengthened by a more rational theoretical approach using stochastic calculus. 

\subsubsection{NEMD results}

\begin{figure} [htbp]
\begin{center}
\includegraphics[width=10cm]{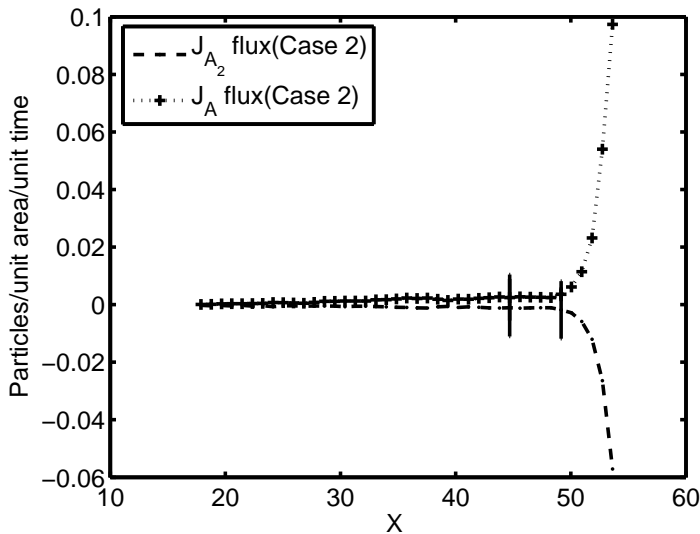}
\caption{Extreme thermodynamical conditions leading to the presence of  steady state atomic and dimer fluxes. 
The data points are used to construct the difference equation in the text to verify the conservation law.  }\label{fig:23}
\end{center}
\end{figure}

\begin{figure} [htbp]
\begin{center}
\includegraphics[width=10cm]{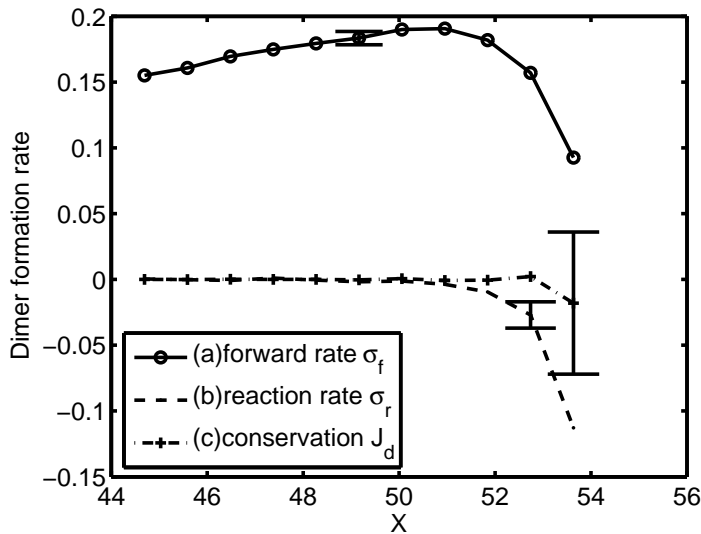}
\caption{Test of divergence theorem for mass conservation via a difference equation.}\label{fig:24}
\end{center}
\end{figure}

 Figs.~(\ref{fig:23},\ref{fig:24}) are the flux and divergence of the flux for "`Case 2"' simulation where a temperature gradient across the MD cell is imposed together with the making and breaking of bonds at the ends of the cell leading to a  molecular flux according to the thermodynamical conditions and details given in \cite{cgj6}. The cell is broken up into 64 layers along the $\text{X}$-direction and the thermostats are placed at the ends of the layers. Fig.~(\ref{fig:23})has overlapping error bars with magnitudes that do not change significantly over the range where the fluxes are evident.  The stationary source and sink quantities are denoted $\sigma$ ($\sigma_f$ and  $\sigma_b$ are the rate of formation and breakdown of the dimer in unit time and unit volume respectively throughout the cell. The conservation of mass equation for atoms and dimers read as follows, where the subscripts refer to the species label for the flow vector $J$ and the concentration $c$:
 
\begin{eqnarray} \nonumber
dc_{A_2 } /dt  &=&  - \nabla .J_{A_2 }  + \sigma _f  - \sigma _b \\ 
dc_A /dt &=&  - \nabla .J_A  - 2\sigma _f  + 2\sigma _b 
\end{eqnarray}
 
The steady state condition is $ \nabla .J_A  =  - 2(\sigma _f  - \sigma _b ) =  - 2\sigma _r 
$ and $\nabla .J_{A_2 }  = \sigma _r $ ; $ (\sigma _f  - \sigma _b ) = \sigma _r $ where $\sigma _r $
is a scalar flux and at thermodynamical equilibrium, $\sigma _r = 0$ strictly. If the PLE were strictly valid then 
the $J_A ,J_{A_2 } $ fluxes must vanish; clearly here, this is not the case. To check for flux conservation, the divergence term is discretized by integration over one layer, using the trapezoidal rule, where 
for any  layer  $i$,
\begin{equation}
\int_{i - 1}^i {\nabla .J_{A_2 } } dV = \frac{{(\sigma _r (i) - \sigma _r (i - 1))\Delta V}}{2} = J_{A_2 ,dif} (i) = J_{A_2 } (i) - J_{A_2 } (i - 1)
\end{equation} 
where the layer has volume $\Delta V$. Similarly, for the atomic fluxes,
\begin{equation}
J_{A,dif} (i) = J_A (i) - J_A (i - 1) =  - (\sigma _r (i) + \sigma _r (i - 1))\Delta V
\end{equation}
leads to 
\begin{equation}\label{e:33}
J_d (i) = 2J_{A_2 ,dif} (i) + J_{A,dif} (i) = 0
\end{equation}
The plot of $J_d $ given in (\ref{e:33})in Fig.~(\ref{fig:24})complies with the conservation law rather well, within statistical error. We have therefore shown that PLE is not a rigorous principle from numerical simulation where a counter-example is given, and that local  stochastic equilibrium dynamical variables  do not necessarily have Gaussian (Canonical) distributions as demanded by some specialists \cite{rubi2,rubi3,kei2} in their theories.
 
\section{Conclusion}

This study shows that the model of the molecule utilizing switching potentials does lead to typical behavior predicted from standard theories for unusual hysteresis-type reaction mechanisms which theorists have largely ignored, due perhaps to the influence of "`time-reversible"'symmetry concepts. It is demonstrated that microscopic loop-like pathways does not influence the macroscopic thermodynamical results in any fundamental way. The method used here to reduce expensive 3-body calculations to easier 2-body calculations  may be used as a basis for non-equilibrium simulation  applications, which will be the subject of further investigations.  The   two body potentials yield extremely  good thermodynamic results whilst being super-efficient in reducing computational costs because the use of switches and algorithms that can preserve momentum and energy during potential transitions, and it is expected that semi-quantitative results at least can be determined for any molecular potential that is known. The NEWAL algorithm is  effective for the extreme conditions of the simulation, and would prove to be a valuable tool in reducing errors attributable to switching potentials. A whole generation of scientific literature has been devoted to establishing necessary connections between the direction of material flow (microscopic reversibility or "`time reversibility"')  and thermodynamics, but the results here suggests that there need not be any necessary connection between the two.  Lastly, it is shown through counter-examples that the PLE and the canonical averaging assumption used in recent thermodynamical theories are not strictly  correct since internal variables do not have the same algebraic structure as the variables that are explicitly featured in the system Hamiltonian. We have demonstrated that there is a feasibility of developing an extended theory of equipartition (where the temperature parameters associated with any  species motion need not be fixed and of the same value as the system thermodynamical temperature) on the basis of the shape of the energy distributions. It would be of interest to repeat and compare some of the above calculations for a conventional system without hysteresis to rule out any  necessary connection between dynamics and equilibrium thermodynamic properties.  
 
\textbf{Acknowledgement}
C.G.J would like to thank (a) University of Malaya, Kuala Lumpur for financing a sabbatical visit to NTNU (2000-2001), and (b) my hosts  S.K. and B.H. of the Institute of Physical Chemistry, NTNU during this period.
\bibliography{ed05}
\end{document}